\documentclass[11pt, a4paper]{article}
\usepackage{showlabels}
\usepackage{graphicx}
\usepackage{epstopdf}
\usepackage{cite}
\usepackage[multiple]{footmisc}

\usepackage{color}
\usepackage{bm}
\usepackage{slashed}
\usepackage{xcolor} 
\usepackage{graphicx}
\usepackage{amsmath}
\usepackage{amssymb}
\usepackage{amsfonts}
\usepackage{amsthm}
\usepackage{mathrsfs}
\usepackage{array}

\usepackage{epsfig}

\usepackage[
      colorlinks=true,
      linkcolor=blue,
      urlcolor=blue,
      filecolor=blue,
      citecolor=red,
      pdfstartview=FitV,
      pdftitle={},
      pdfauthor={},
      pdfsubject={},
      pdfkeywords={},
      pdfpagemode=None,
      bookmarksopen=true
]{hyperref}

\usepackage{epsfig}

\usepackage{hyperref}
\let\oldFootnote\footnote
\newcommand\nextToken\relax

\renewcommand\footnote[1]{%
    \oldFootnote{#1}\futurelet\nextToken\isFootnote}

\newcommand\isFootnote{%
    \ifx\footnote\nextToken\textsuperscript{,}\fi}

\def\beq{\begin{eqnarray}}
\def\eeq{\end{eqnarray}}

\def\be{\begin{equation}}
\def\ee{\end{equation}}
\def\bea{\begin{eqnarray}}
\def\eea{\end{eqnarray}}

\def\be{\begin{equation}}
\def\ee{\end{equation}}
\def\bea{\begin{eqnarray}}
\def\eea{\end{eqnarray}}

\newcommand{\rom}[1]{\mathrm{#1}}

\def\cA{\mathcal{A}}

\def\cF{\mathcal{F}}

\def\nn{\nonumber}

\numberwithin{equation}{section}

\begin{document}

\begin{centering}

\textbf{\LARGE{Horizon Hair from Inversion Symmetry}}

 \vspace{0.8cm}
Karan Fernandes$^{1,2}$, Debodirna Ghosh$^2$, 
Amitabh Virmani$^2$

  \vspace{0.5cm}

\begin{minipage}{.9\textwidth}\small  \begin{center}
$^1$Harish-Chandra Research Institute, Chhatnag Road, \\
Jhusi, Allahabad, India 211019  \\ \vspace{0.5cm}
{\tt  karanfernandes@hri.res.in}
  \vspace{0.5cm}
  \\ 
$^2$Chennai Mathematical Institute, H1 SIPCOT IT Park, \\ Kelambakkam, Tamil Nadu, India 603103\\
  \vspace{0.5cm}
{\tt  debodirna, avirmani@cmi.ac.in}

\end{center}
\end{minipage}

 \vspace{0.5cm}

\end{centering}

\begin{abstract} 
The extreme Reissner-Nordstr\"om solution has a discrete conformal isometry that maps the future event horizon to future null infinity and vice versa, the Couch-Torrence inversion isometry. We study the dynamics of a probe Maxwell field on the extreme Reissner-Nordstr\"om solution in light of this symmetry. We present a gauge fixing that is compatible with the inversion symmetry. The gauge fixing allows us to relate the gauge parameter at the future horizon to future null infinity, which further allows us to study global charges for large gauge symmetries in the exterior of the extreme Reissner-Nordstr\"om black hole. Along the way, we construct Newman-Penrose and Aretakis like conserved quantities along future null infinity and the future event horizon, respectively, and relate them via the Couch-Torrence inversion symmetry.  
\end{abstract}

 \vspace{1.5cm}
 \begin{center}
{\textit{Dedicated to the memory of Prof.~Pushan Majumdar, 1972-2020.}}
\end{center}

\newpage

\tableofcontents

\section{Introduction}
In general relativity, diffeomorphisms that preserve fall-off conditions near null infinity give rise to the infinite-dimensional BMS group \cite{Bondi:1962px, Sachs:1962wk, Sachs:1962zza}. In recent years, it has been shown that there are closely related infinite-dimensional symmetries consisting of large gauge transformations for quantum electrodynamics (QED) \cite{He:2014cra, Kapec:2015ena} in Minkowski spacetime.  The gauge parameter is an arbitrary function on the sphere
$\epsilon_{\mathcal{I}^+} (z, \bar z)$ at future null infinity. These symmetries enable one to view soft photon theorems in QED as associated Ward identities \cite{1703.05448}.

Related developments have found that stationary black holes also possess an infinite number of symmetries in the near horizon region \cite{Koga:2001vq, Donnay:2015abr, Mao:2016pwq, Donnay:2016ejv,  Hawking:2016msc, Hawking:2016sgy, Carlip:2017xne, Blau:2015nee, Penna:2017bdn, Grumiller:2018scv, Chandrasekaran:2018aop, Donnay:2018ckb, Chen:2020nyh, Perry:2020ndy, Donnay:2020yxw}\footnote{The references are representative of the very large literature on the subject.}\footnote{The symmetry groups in the referenced papers do not necessarily coincide. This is so because different authors preserve different structures: some prefer to preserve a particular geometric structure on the null surface, whereas others preserve the near horizon geometry. Reference \cite{Donnay:2018ckb} performs the near-horizon asymptotic symmetry analysis for the Reissner-Nordstr\"om black hole in the Einstein-Maxwell theory.}. Often, the symmetries are diffeomorphisms that preserve a notion of the  near horizon geometry or diffeomorphisms that preserve a particular geometric structure on the horizon. Typically, a class of these symmetries is similar to supertranslations at null infinity.  It is believed that global charges associated with supertranslations receive contributions from the  horizon as well as from null infinity.  A complete discussion of conservation laws associated with supertranslations requires a detailed understanding of how the symmetries at the horizon relate to the symmetries at null infinity. However, the precise relation between the horizon and null infinity symmetries has not been sufficiently understood. It is therefore of considerable interest to understand, say, even in a toy model, the relation between the horizon and null infinity symmetries. The aim of this work is to make progress on this issue in the context of the dynamics of a probe Maxwell field on the extreme Reissner-Nordstr\"om (ERN) black hole spacetime. 

To some extent these issues were explored in \cite{Hawking:2016msc, 1703.05448}, where electromagnetic soft-hair shockwaves into the Schwarzschild black hole were considered.\footnote{In a gravitational setting, 
references~\cite{Hawking:2016sgy, Compere:2016hzt} consider throwing a soft-hair shockwave into the Schwarzschild black hole; reference \cite{Donnay:2018ckb} considers soft-hair shockwaves in the Reissner-Nordstr\"om black hole.}  In these references, the gauge $A_v=0$ in advanced Bondi coordinates was used.
This gauge is natural in analysing how excitations from past null infinity relate to excitations near the future horizon. However, since the advanced Bondi coordinates do not cover future null infinity, the relation between gauge parameters at future null infinity and the future horizon remains unexplored (at best indirect).\footnote{The importance of understanding this relation is emphasised by Chandrasekaran-Flanagan-Prabhu in \cite{Chandrasekaran:2018aop}.}

The aim of this work is to  overcome this shortcoming in a toy model. We hope that the fundamental ideas will find broader applicability. Our toy model is the dynamics of a probe Maxwell field on the exterior of the ERN black hole.
The ERN background enjoys a \emph{discrete conformal} symmetry. The symmetry acts as a spatial inversion interchanging the future event horizon $\mathcal{H}^+$   and  future null infinity $\mathcal{I}^+$: the Couch-Torrence (CT) inversion symmetry \cite{ct:1984}.\footnote{The CT inversion also interchanges the past event horizon $\mathcal{H}^-$ and past null infinity $\mathcal{I}^-$. However, for most of the paper we will be only concerned with the mapping between the future event horizon and future null infinity.} This inversion symmetry is the key property of the ERN spacetime used in this work.

The organisation of the  rest of the paper is as follows. 
 The  CT symmetry and its action on a probe Maxwell field is presented in section \ref{sec:CT}. In section \ref{sec:ES_gauge} we present a study of a CT invariant gauge fixing for the probe Maxwell field on the ERN background. We show that the CT invariant gauge fixing is closely related to the harmonic gauge in an asymptotic expansion near null infinity. Specifically, we show that in an asymptotic expansion, the harmonic gauge condition is compatible with the CT invariant gauge condition. We then analyse the CT invariant gauge condition near the horizon in an asymptotic expansion. We conclude that at the future horizon too, the gauge parameter is an arbitrary function on the sphere $\epsilon_{\mathcal{H}^+}
 (z, \bar z)$ independent of the ingoing Eddington-Finkelstein coordinate $v$.  
 
 \begin{figure}[t]
\begin{center}
 \includegraphics[width=0.95\textwidth]{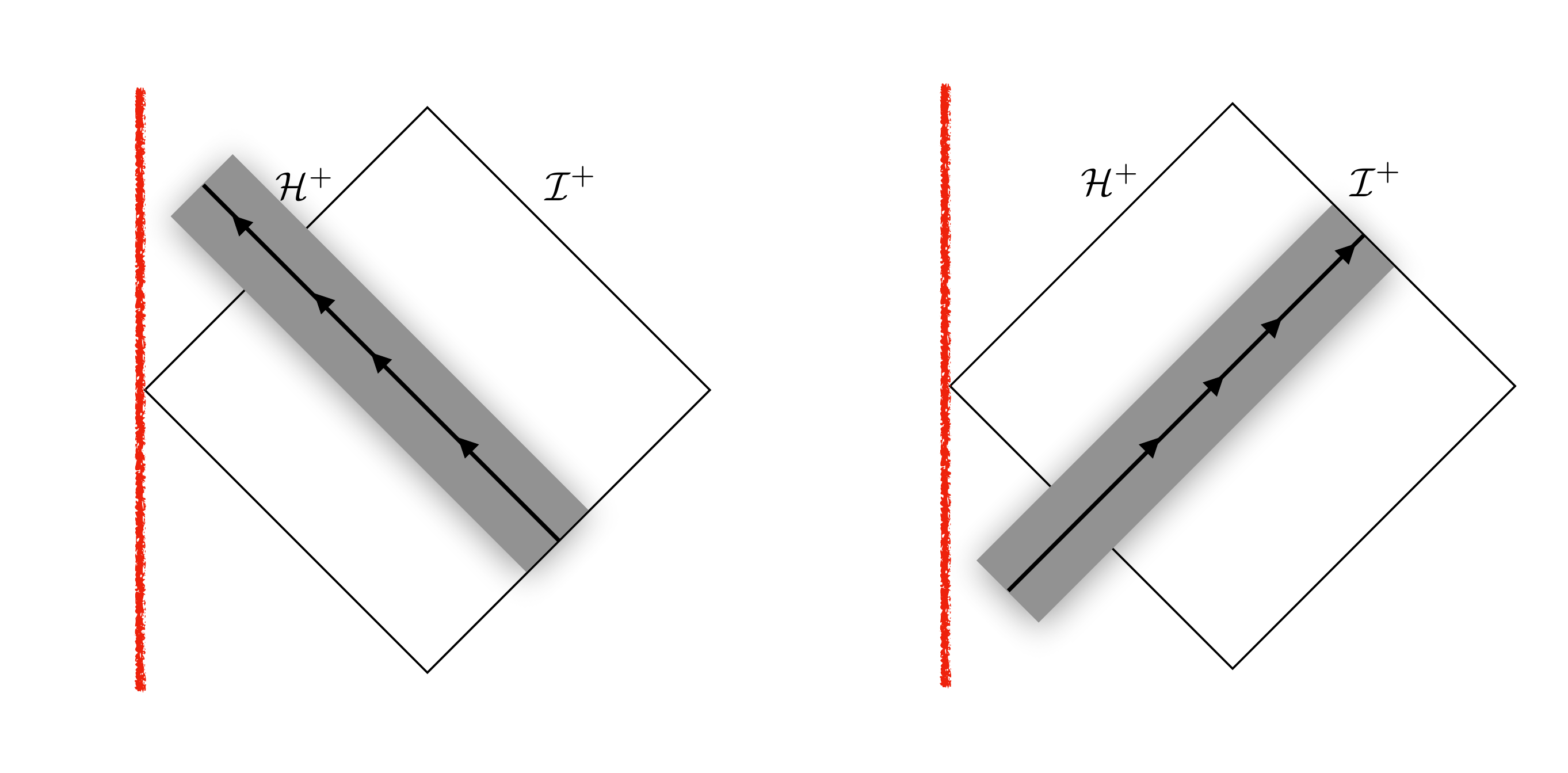}
 \caption{\sf Under the Couch-Torrence (CT)  mapping, a purely ingoing configuration of a probe Maxwell field on the extreme Reissner-Nordstr\"om (ERN) background (left) is transformed to a purely outgoing configuration (right) and vice versa. Later in the paper we show that if the right configuration carries soft charges at future null infinity, the ingoing left configuration carries soft charges at the horizon.}
\label{Fig:CT_mapping}
\end{center}
\end{figure}

How are the two functions $\epsilon_{\mathcal{H}^+}
 (z, \bar z)$ and $\epsilon_{\mathcal{I}^+}
 (z, \bar z)$ related to each other? 
The CT invariant gauge condition leads to  a fourth order differential equation for the residual gauge parameter. The fourth order equation is difficult to analyse.  Motivated by the results of the previous sections,  namely the usefulness of the harmonic gauge and conformal transformations,  we solve a simpler problem that captures the essential ideas in section \ref{sec:toy_model}. We study the residual gauge parameter in the harmonic gauge on a spacetime obtained by a conformal rescaling of the ERN spacetime. This spacetime has two asymptotically flat ends (two future null infinities): one representing the future null infinity of the ERN spacetime and the other representing the horizon. The spacetime has the inversion symmetry. We show that the gauge parameter smoothly extends from an arbitrary function on the sphere from one null infinity to the other null infinity in such a way that 
 $\epsilon_{\mathcal{H}^+}
 (z, \bar z)=\epsilon_{\mathcal{I}^+}
 (z, \bar z)$.

In section \ref{sec:horizon_hair} we present an expression for the global charge (often called the Iyer-Wald charge) for the probe Maxwell field on the ERN spacetime. The charge integral is well defined when we push the Cauchy surface to $\mathcal{I}^+ \cup \mathcal{H}^+$. The answer is written as a sum of two terms: one at $\mathcal{I}^+$ and the other at $\mathcal{H}^+$. We argue that soft electric hairs on the horizon of the ERN spacetime follows from the CT transformation of configurations that have soft electric hairs at null infinity, and vice versa. This is schematically shown in figure \ref{Fig:CT_mapping}.

Finally, in section \ref{sec:NP_charges} we construct  Newman-Penrose and Aretakis like conserved quantities along future null infinity and the future event horizon, respectively.  These constants are  related via the CT inversion symmetry. This section is an extension of the corresponding mapping understood for the massless scalar field on the ERN spacetime.

\section{A probe Maxwell field on the extreme RN background}

\label{sec:CT}

The  ERN solution has a discrete conformal symmetry \cite{ct:1984}, which acts as a spatial inversion  interchanging the future event horizon $\mathcal{H}^+$ and future null infinity $\mathcal{I}^+$. The significance of this inversion map for scalar dynamics has been explored by several authors in the physics literature \cite{Blaksley:2007ak, Bizon:2012we, Lucietti:2012xr, Sela:2015vua, Godazgar:2017igz, Bhattacharjee:2018pqb, Cvetic:2018gss} and in the mathematical general relativity literature \cite{Angelopoulos:2018uwb}. In this paper,  we are interested in the significance of this inversion map
on the dynamics of  a probe Maxwell field. We are especially interested in the transformation of the probe Maxwell field under this symmetry.

Let us consider the metric of the  $4$-dimensional ERN spacetime in static coordinates
\begin{equation}
ds^2=-\left(1-\frac{M}{r}\right)^2 dt^2+\left(1-\frac{M}{r}\right)^{-2} dr^2+r^2d\Omega_2^2\label{ern.met},
\end{equation}
where $d\Omega_2^2$ is the line element of the $2$-sphere. Throughout the paper, we will describe the metric of the $2$-sphere in terms of the complex stereographic coordinates $\left(z,\bar{z}\right)$,
\begin{equation}
d\Omega_2^2 = 2 \gamma_{z \bar{z}}dz d\bar{z} = \frac{4}{\left(1 + z \bar{z}\right)^2} dz d\bar{z} \,.
\end{equation}
The ERN metric in static coordinates admits a discrete conformal symmetry under the CT transformation,
\begin{equation}
\mathcal{T} : \left(t\,,r\,,z\,,\bar{z}\right) \to \left(t\,,   M + \frac{M^2}{r - M}\,,z\,,\bar{z}\right).
\label{ern.ct1}
\end{equation}
The pull-back of this transformation on the metric acts as a conformal transformation
\begin{equation}
{\cal{T}}_{*}(g)=\omega^2 g \qquad \mbox{where} \qquad \omega=\frac{M}{r-M}.
\label{ern.ct2}
\end{equation}
The transformation equation~(\ref{ern.ct1}) is an involution, i.e. $\mathcal{T}^2 = 1$.  On the tortoise coordinate $r_*(r)$, defined by, 
\begin{equation}
r_*(r) = r - M - \frac{M^2}{r-M} + 2 M \log\left(\frac{\vert r- M\vert}{M}\right) \,, \label{ern.tort}
\end{equation}
it acts as $\mathcal{T}: r_* \to - r_*$. This, in particular, implies that it interchanges the ingoing  and outgoing   Eddington-Finkelstein  coordinates $v= t + r_*$ and $u = t - r_*$:
\begin{equation}
\mathcal{T}: u \leftrightarrow v.
\end{equation} 
Hence, it can be concluded that the CT transformation through its action as a spatial inversion interchanges the future event horizon $\mathcal{H}^+$ with future null infinity $\mathcal{I}^+$.

Let us now consider a \emph{probe Maxwell field} on the ERN background. The probe field is different from the Maxwell field under which the ERN black hole is charged.  
Let us denote the probe field by ${\cal{A}}_a$, with the corresponding field strength written as
\begin{equation}
{\cal{F}}_{ab}=\partial_a{\cal{A}}_b-\partial_b{\cal{A}}_a.
\end{equation}
The field satisfies the source-free Maxwell equations 
\begin{equation}
g^{ca} \nabla_c {\cal{F}}_{ab} = 0 ,
\label{max.eq}
\end{equation}
where $\nabla_a$ is the covariant derivative. 
Let us now consider a general conformal transformation of the spacetime metric $g_{ab}$
\begin{equation}
g_{ab} \to \tilde{g}_{a b} = \Omega^2 g_{a b}\,,
\end{equation}
with $\Omega$ being the conformal factor. In four spacetime dimensions Maxwell's equations are known to be conformally invariant
\begin{equation}
\tilde \nabla^a{\cal{F}}_{ab}\rightarrow \Omega^{-2}\nabla^a{\cal{F}}_{ab}, \qquad \tilde \nabla_{[a} {\cal{F}}_{bc]} \to \nabla_{[a} {\cal{F}}_{bc]} 
\end{equation}
with conformal weight zero.

Using equation~(\ref{ern.ct2}) in equation~(\ref{max.eq}), it follows that,
\begin{equation}
0=\nabla^a{\cal{F}}_{ab}={\cal{T}}_*(\nabla^a{\cal{F}}_{ab})=\nabla^a_{{\cal{T}}_*(g)}({\cal{T}}_*{\cal{F}}_{ab})=\nabla^a_{\omega^2 g}({\cal{T}}_*{\cal{F}}_{ab})=\omega^{-2}(\nabla^a({\cal{T}}_*{\cal{F}}_{ab})).
\label{max.ct2}
\end{equation}
Thus, if ${\cal{F}}_{ab}$ is a solution of  Maxwell’s equations, then so is ${\cal{T}}_*{\cal{F}}_{ab}$. Specifically, in coordinates, if ${\cal{F}}_{ab} (x) $
is a given solution in static coordinates $x^a = \{t, r, z, \bar{z}\}$, then 
\be
({\cal{T}}_*{\cal{F}}_{ab}) (y) = \left( \frac{\partial x^c }{\partial y^{a}} \right) \left(  \frac{\partial x^d }{\partial y^{b}}\right)  
{\cal{F}}_{cd} (x), 
\ee
is a \emph{different} solution in static coordinates where $y^{a}=\left\{t,  M + \frac{M^2}{r - M}, z, \bar{z}\right\} .$
Since the coordinate transformation only changes the radial coordinate, only the radial components of the two-form field pick up additional factors. That is, if ${\cal{F}}_{ir}$ and ${\cal{F}}_{ij}$ for $i=t, z, \bar{z}$ is a solution, then  so is
\bea
{\cal{F}}_{ir} (t, r, z, \bar{z})&=& - \frac{M^2}{(r-M)^2} {\cal{F}}_{ir}\left(t, M + \frac{M^2}{r - M}, z, \bar{z}\right), \\
{\cal{F}}_{ij} (t, r, z, \bar{z}) &=&  {\cal{F}}_{ij}\left(t, M + \frac{M^2}{r - M}, z, \bar{z}\right).
\eea

In terms of ingoing and outgoing coordinates, if ${\cal{F}}_{vr}(v, r, z, \bar{z})$ and ${\cal{F}}_{z\bar{z}}(v, r, z, \bar{z})$ are components of a given solution in the \emph{ingoing} Eddington-Finkelstein coordinates then, 
\be
{\cal{F}}_{ur} (u, r, z, \bar{z})= -\frac{M^2}{(r-M)^2} {\cal{F}}_{vr}\left(u, M + \frac{M^2}{r - M}, z, \bar{z}\right),
\label{fur.ct}
\ee
and
\be
{\cal{F}}_{z\bar{z}}(u, r, z, \bar{z})= {\cal{F}}_{z\bar{z}}\left(u, M + \frac{M^2}{r - M}, z, \bar{z}\right),
\ee
are the components of a \emph{different} solution in \emph{outgoing} Eddington-Finkelstein coordinates.

\section{Eastwood-Singer Couch-Torrence invariant gauge condition}
\label{sec:ES_gauge}

While solutions of Maxwell's equations are invariant under conformal transformations, this property does not extend to arbitrary gauge fixings of the Maxwell field. For instance,  as is well known the harmonic gauge $\nabla^a \mathcal{A}_a =0$ is not conformally invariant~\cite{Cote:2019kbg}. 
 Under conformal transformations $\tilde g_{ab} = \Omega^2 g_{ab}$, 
\begin{equation}
\nabla^a \mathcal{A}_a \to \widetilde{\nabla}^a \tilde{\mathcal{A}}_a = \Omega^{-2}\left(\nabla^a\mathcal{A}_a + 2 \Upsilon^a\mathcal{A}_a\right), \label{es.lorc}
\end{equation}
where $ \Upsilon_a = \nabla_a \ln \Omega\,.$ Tildes will be used to denote all conformally transformed objects.

A conformally invariant gauge choice for the source-free Maxwell equations was introduced by Eastwood and Singer~\cite{Eastwood:1985eh}:
\begin{equation}
\mathcal{D}^a\mathcal{A}_a := \nabla_b\left(\nabla^b \nabla^a - S^{ab}\right)\mathcal{A}_a = 0 \,,
\label{es.eq}
\end{equation}
where
\begin{equation}
S^{ab} = - 2 R^{ab} + \frac{2}{3}R g^{ab}.
\label{es.s}
\end{equation}

The invariance of  equation~(\ref{es.eq})  under conformal transformations  for the source-free Maxwell equations can be established via a straightforward, if somewhat, tedious calculation. Consider  $\nabla^b \nabla^a {\cal{A}}_a$ and $S^{ab}$ terms separately. 
The conformal transformation of the tensor $S^{ab}$ defined in equation~(\ref{es.s}) is:
\begin{equation}
\tilde{S}^{ab} = \Omega^{-4}\left(S^{ab} + 2 \nabla^{(a}\Upsilon^{b)} -2 g^{ab}\nabla^c\Upsilon_c - 4 \Upsilon^a\Upsilon^b \right) \,.
\label{es.Sc}
\end{equation}
The conformal transformation of the term $\nabla^b \nabla^a {\cal{A}}_a$ is:
\begin{equation}
\tilde{\nabla}^b\tilde{\nabla}^a\tilde{{\cal{A}}}_a = \tilde{\nabla}^b\left(\Omega^{-2}\left(\nabla^a + 2 \Upsilon^a\right)\tilde{{\cal{A}}}_a\right) = \Omega^{-4}\left(\nabla^b - 2 \Upsilon^b\right)\left(\nabla^a + 2 \Upsilon^a\right){\cal{A}}_a
\label{es.ddac}
\end{equation}
Combining the above two results, the transformation of $\mathcal{D}^a\mathcal{A}_a $ is,
\begin{equation}
\tilde{\mathcal{D}}^a \tilde{\mathcal{A}}_a = \tilde{\nabla}_b\left(\tilde{\nabla}^b\tilde{\nabla}^a - \tilde{S}^{ab}\right)\tilde{{\cal{A}}}_a = \tilde{\nabla}_b\left(\Omega^{-4}\mathcal{V}^b\right),
\label{es.dac}
\end{equation}
where
\begin{equation}
\mathcal{V}^b = \left(\nabla^b\nabla^a - S^{ab}\right){\cal{A}}_a + 2 \left[ \left(\Upsilon^{a} \nabla^{b} - \Upsilon^{b} \nabla^{a}\right) \mathcal{A}_a + \mathcal{A}^b \nabla_a \Upsilon^a - \mathcal{A}_a\nabla^a \Upsilon^b\right].
\end{equation}
Next we observe that for an arbitrary vector $V^a$, 
\be
\tilde{\nabla}_b\left(\Omega^{-4} V^{b}\right) = \Omega^{-4}\nabla_b V^b \label{es.id2}.
\ee
To demonstrate the conformal invariance of the Eastwood-Singer gauge, we use equation~(\ref{es.id2}) in the last expression of equation~(\ref{es.dac}) to find
\begin{align}
 \tilde{\mathcal{D}}^a \tilde{\mathcal{A}}_a = \Omega^{-4} \nabla_a\mathcal{V}^a 
= \Omega^{-4} \left(\mathcal{D}^a \mathcal{A}_a + 2 \Upsilon^b \nabla^a \mathcal{F}_{ab}\right)
= \Omega^{-4} \mathcal{D}^a \mathcal{A}_a.
\end{align} 
In arriving at the last equality, we made use of the source-free Maxwell equations  $\nabla^a \mathcal{F}_{ab}=0$. Thus equation~(\ref{es.eq}) is invariant under conformal transformations.

This gauge condition is also invariant under the CT transformation on the ERN spacetime. Denoting the CT transformed gauge field as $\mathcal{T}_*{\mathcal{A}}_a$, we have,  
\begin{equation}
 {\cal{T}}_*\left(\mathcal{D}^a \mathcal{A}_a\right) =  \nabla^{{\cal{T}}_*(g)}_b\left(\nabla_{{\cal{T}}_*(g)}^b\nabla_{{\cal{T}}_*(g)}^a - {\cal{T}}_*(S^{ab})\right){\cal{T}}_*{\cal{A}}_a = \omega^{-4} \mathcal{D}^a \left({\cal{T}}_*{\cal{A}}_a\right) = 0,
\end{equation}
where we have used $\mathcal{T}_*(g) = \omega^2 g$ from equation~(\ref{ern.ct2}).
Hence, if ${\cal{A}}_a$ is gauge fixed by equation~(\ref{es.eq}), then the CT transformed gauge field ${\cal{T}}_*{\cal{A}}_a$ also satisfies the same gauge condition. 

We are now in position to discuss the residual gauge transformations for the Eastwood-Singer gauge. Under the gauge transformations $\mathcal{A}_a \to \mathcal{A}_a + \nabla_a \epsilon$, the expression in (\ref{es.eq}) provides the equation satisfied by the residual gauge  parameter $\epsilon$, 
\begin{equation}
\mathcal{D}^a\nabla_a \epsilon = \nabla_b\left(\nabla^b \Box + \left(2 R^{ab} - \frac{2}{3}R g^{ab}\right)\nabla_a\right) \epsilon = 0\,,
\label{es.rgf} 
\end{equation}
where $\Box = g^{ab}\nabla_a \nabla_b = g^{ab}\left(\partial_a\partial_b - \Gamma_{ab}^c\partial_c\right)$ is the D'Alembertian operator on curved spacetimes. Equation \eqref{es.rgf} simplifies on the ERN spacetime, for which $R=0$ and $\nabla_a R^{ab} = 0$ (from Einstein's equations). As a result, equation~(\ref{es.rgf}) on the ERN spacetime becomes
\begin{align}
\left(\Box g^{ab} + 2 R^{ab}\right)\nabla_a \nabla_b \epsilon= \Box \Box \epsilon + 2 R^{ab}\partial_a\partial_b \epsilon - 2 R^{ab}\Gamma^c_{ab} \partial_c \epsilon =0.
\label{es.reRN}
\end{align}
This is a fourth-order equation for $\epsilon$ in the exterior of the ERN spacetime. Our interest in residual gauge transformations is largely in the context of soft charges at the asymptotic boundaries of the spacetime, namely at future null infinity $\mathcal{I}^+$ and the future event horizon $\mathcal{H}^+$. We also note that if $\epsilon$ is a function satisfying \eqref{es.reRN} then so is ${\cal{T}}_* \epsilon$.

To investigate possible solutions for  $\epsilon$ near $\mathcal{I}^+$ we use the ERN metric in outgoing Eddington-Finkelstein coordinates
\begin{equation}
ds^2 = -\left(1 - \frac{M}{r}\right)^2 du^2 - 2 du dr + r^2d\Omega_2^2. \label{es.oef}
\end{equation}
 Let us quickly recall the discussion for the harmonic gauge. Inserting the ansatz \cite{1703.05448},
\bea
\epsilon(r, u, z, \bar{z}) &=& \epsilon^{(0)} (u,z,\bar z) + \frac{1}{r} \epsilon^{(1)} (u,z,\bar z)+  \frac{1}{r} f^{(1)}(u,z,\bar z) \log \frac{u}{2r} \nn \\ & & \quad \quad + \frac{1}{r^2} \epsilon^{(2)}(u,z,\bar z) +  \frac{1}{r^2} f^{(2)}(u,z,\bar z) \log \frac{u}{2r} + \mathcal{O}(r^{-3}),
\label{gp.inf}
\eea
in the scalar equation 
$
\Box \epsilon = 0,
$
and expanding in powers of large $r$, we find the following equations order-by-order in inverse powers of $r$,
\bea
\partial_u \epsilon^{(0)}  &= & 0, \label{gf.1} \\
\partial_u f^{(1)}  &= & - \frac{1}{2} D^2 \epsilon^{(0)}, \\
\partial_u f^{(2)}  &= & - \frac{1}{2} D^2 f^{(1)}, \\
\partial_u \epsilon^{(2)}  &= & \frac{1}{2} D^2 f^{(1)}  - \frac{1}{2} D^2 \epsilon^{(1)} - \frac{1}{2} f^{(1)} - \frac{1}{u}f^{(2)}.\label{gf.4}
\eea
The first of these equations tells us that $\epsilon^{(0)} (u,z,\bar z) $ is independent of $u$: $\epsilon^{(0)} (u,z,\bar z) =: \epsilon_{\mathcal{I}^+}(z, \bar z)$.

Using expansion \eqref{gp.inf} and equations~\eqref{gf.1}--\eqref{gf.4}, a calculation shows that the gauge condition \eqref{es.reRN} is satisfied in an expansion in inverse powers of $r$. Let us demonstrate how this works.  Inserting expansion \eqref{gp.inf} in equation~\eqref{es.reRN}, the leading order term is  $\mathcal{O}(r^{-3})$, whereas the last two terms of equation~\eqref{es.reRN} start at $\mathcal{O}(r^{-6})$ and $\mathcal{O}(r^{-5})$ respectively. Hence, the last two terms do not contribute at order $\mathcal{O}(r^{-3})$ and $\mathcal{O}(r^{-4})$.  As a result, at these orders, the gauge fixing equation simply becomes 
\be
 \Box \Box \epsilon  = 0. \label{boxbox}
\ee
At  order  $\mathcal{O}(r^{-3})$  equation \eqref{boxbox} gives,
\be
\partial_u^2 f^{(1)} = 0, \label{first_eq_new_gauge_u}
\ee
which is consistent with equations \eqref{gf.1}--\eqref{gf.4}, in this sense that if those equations are satisfied then  \eqref{first_eq_new_gauge_u} is also satisfied: 
\be
\partial_u (\partial_u f^{(1)}) =  \partial_u \left( - \frac{1}{2} D^2 ( \epsilon^{(0)}) \right) = - \frac{1}{2} D^2 ( \partial_u \epsilon^{(0)}) = 0.
\ee
At order $\mathcal{O}\left(r^{-4}  \log \frac{u}{2r}\right) $ equation \eqref{es.reRN}  or \eqref{boxbox}   gives, 
\be
\partial_u^2 f^{(2)} +  \frac{1}{2} D^2 ( \partial_u f^{(1)}) =0.
\ee
This is also consistent with equations \eqref{gf.1}--\eqref{gf.4} as,
\be
\partial_u^2 f^{(2)} +  \frac{1}{2} D^2 (\partial_u f^{(1)}) = \partial_u \left(\partial_u f^{(2)} + \frac{1}{2} D^2 f^{(1)}\right) = 0.
\ee
At  order  $\mathcal{O}\left(r^{-4}\right)$ the details are a little more cumbersome. One finds, 
\bea
&& D^2 D^2 \epsilon^{(0)} + 2 D^2 \epsilon^{(0)} + 4 M \partial_u \epsilon^{(0)}  + \frac{4}{u} D^2 f^{(1)} + 8 \partial_u f^{(1)}   + 4 D^2 (\partial_u f^{(1)})  + 4 D^2 (\partial_u \epsilon^{(1)})   \nonumber \\
&&  12 \partial_u^2f^{(2)}  + \frac{16}{u} \partial_u f^{(2)} - \frac{8}{u^2}  f^{(2)} + 8 \partial_u^2 \epsilon^{(2)} =0. \label{cumbersome_u}
\eea
Again, one can check that this equation is consistent with equations \eqref{gf.1}--\eqref{gf.4}.\footnote{The easiest way to confirm this is to substitute $\partial_u \epsilon^{(2)}$ in \eqref{cumbersome_u} from \eqref{gf.4}.}

As  argued above, if $\epsilon $ is a solution to the gauge condition then so is $\mathcal{T}_* \epsilon $. The expansion of
\be
 \epsilon (r, v, z, \bar z)= \mathcal{T}_* ( \epsilon(r, u,z,\bar z)),
\ee
 near the horizon takes the form, cf.~\eqref{ern.ct1}, 
\bea
 && \tilde \epsilon(r, v, z, \bar{z}) = \tilde \epsilon^{(0)} (v,z,\bar z) 
  + \frac{(r-  M)}{r M} \left\{ \tilde \epsilon^{(1)} (v,z,\bar z)+  \tilde f^{(1)}(v,z,\bar z) \log \left[  \frac{v(r-  M)}{2r M}  \right] \right\} \nn \\ & & \quad \quad + \frac{(r-  M)^2}{r^2 M^2} \left\{ \tilde \epsilon^{(2)}(v,z,\bar z) +   \tilde  f^{(2)}(v,z,\bar z) \left[ \frac{v(r-  M)}{2 r M} \right] \right\}    + \mathcal{O}((r-M)^{3}).
\label{gp.horizon}
\eea
Using ingoing Eddington-Finkelstein coordinates, 
\begin{equation}
ds^2 = -\left(1 - \frac{M}{r}\right)^2 dv^2 + 2 dv dr + r^2d\Omega_2^2, \label{es.ief}
\end{equation}
and substituting the series expansion \eqref{gp.horizon} in gauge condition \eqref{es.reRN} we find that the first non-trivial term appears at order $(r-M)^{-1}$. The conditions at orders $(r-M)^{-1}$, $\log \left[ \frac{v(r-  M)}{2 r M} \right] $, and $\mathcal{O}(1)$ respectively give, 
\be
\partial_v^2 \tilde f^{(1)} = 0, \label{first_eq_new_gauge}
\ee
\be
\partial_v^2 \tilde f^{(2)} + \frac{1}{2} D^2 (  \partial_v \tilde f^{(1)}) =0,
\ee
and
\bea
&& D^2 D^2 \tilde \epsilon^{(0)} + 2 D^2 \tilde \epsilon^{(0)} + 4 M \partial_v \tilde \epsilon^{(0)}  + \frac{4}{v} D^2 \tilde f^{(1)} + 8 \partial_v \tilde f^{(1)}   + 4 D^2 (\partial_v \tilde f^{(1)})  + 4 D^2 (\partial_v \tilde \epsilon^{(1)})   \nonumber \\
&&  12 \partial_v^2 \tilde f^{(2)}  + \frac{16}{v} \partial_v \tilde f^{(2)} - \frac{8}{v^2}  \tilde f^{(2)} + 8 \partial_v^2 \tilde \epsilon^{(2)} =0. \label{cumbersome}
\eea
These equations are identical  to the ones obtained earlier in the expansion near null infinity. Thus, we conclude that tilde variables satisfying equations \eqref{gf.1}--\eqref{gf.4} (with $u$ replaced with $v$ at all places) determine the gauge parameter for $\tilde \epsilon(r, v, z, \bar{z}) $ near the horizon as well. At the horizon  $\lim_{r\to M} \tilde \epsilon(r, v, z, \bar{z}) = \tilde \epsilon^{(0)} (z,\bar z) =: \epsilon_{\mathcal{H}^+}(z, \bar z) $: an arbitrary function on the sphere. 

To summarise: near null infinity we take the ansatz \eqref{gp.inf} for the gauge parameter. The various functions entering the expansion are taken to  satisfy equations~\eqref{gf.1}--\eqref{gf.4}, which are those for the harmonic gauge choice at null infinity. Such a choice is consistent with the Eastwood-Singer gauge fixing~\eqref{es.reRN}. Next, using the CT inversion symmetry of the Eastwood-Singer gauge fixing condition, we obtain expansion  \eqref{gp.horizon} near the horizon. The same functions enter the expansion as near null infinity, except that at all places $u$ is replaced by $v$.   We conclude that  CT inversion allows us to consider gauge fixing such that at the future horizon too, the gauge parameter is an an arbitrary function on the sphere independent of $v$. At this stage the two arbitrary functions on the sphere $\epsilon_{\mathcal{H}^+}(z, \bar {z})$  and $\epsilon_{\mathcal{I}^+}(z, \bar {z})$ are not related. In next section we conjecture that it is natural to expect that they are the same.

It is an important question to explore how the above solutions for the gauge parameter are consistent with the decay results for scalars in black hole spacetimes.  Equally important is to explore the asymptotic dynamics of Maxwell's equation in terms of the gauge field $\cA_a$ in the Eastwood-Singer gauge on the ERN spacetime. We leave these questions for future work. 

Note that the gauge parameter in the Eastwood-Singer gauge has conformal weight zero. This is related to it satisfying a specific fourth order differential equation. The gauge parameter in the harmonic gauge on the other hand satisfies a second order differential equation. If we require this equation to be conformally invariant, then the conformal weight of the scalar  should be  one.

\section{Hyperboloidal slicing and a toy model problem}
\label{sec:toy_model}
In the previous section we showed that large gauge transformations on ERN spacetime consists of the union of two types of transformations. First, where the gauge parameter is an arbitrary function on the sphere independent of $v$ at the horizon. Second, where the gauge parameter is an  arbitrary function on the sphere independent of $u$ at null infinity. The two sets are related by the CT inversion symmetry. It is natural to ask if there exists a smooth interpolation between the two.

The Eastwood-Singer residual gauge parameter  satisfies a fourth order differential equation. It is difficult to analyse that equation. As we saw in the previous section, in an expansion near null infinity the Eastwood-Singer gauge condition is compatible with the harmonic gauge condition.
For this reason, in this section, we study  the residual gauge parameter in the harmonic gauge   on a spacetime conformal related to ERN spacetime as a toy model problem.  We do so in a convenient `hyperboloidal' slicing and show that  $ \epsilon_{\mathcal{H}^+}(z, \bar {z}) = \epsilon_{\mathcal{I}^+}(z, \bar {z})$ .  We conjecture that the same is true for the Eastwood-Singer gauge parameter in the ERN spacetime. 

The hyperboloidal slices \cite{Bizon:2016flx, Zenginoglu:2007jw} parameterised by coordinate $s$  (introduced below) intersect future null infinity along an outgoing null line  at the retarded time $u=s$. They intersect the future horizon at the advanced time $v=s$, however, the normal to the $s=$ constant surfaces at the future horizon is not null; it is timelike. The key point being that the slices intersect both the future null infinity and the future horizon.

To set up these coordinates, we begin by introducing,
 \be
 \tilde t  = \frac{t}{4M} \qquad \mbox{and} \qquad x = \ln \left( \frac{r}{M} - 1\right),
 \ee
 in terms of which the metric takes the form,
 \be
 ds^2 = \frac{16 M^2}{(1+ e^{-x})^2} \left( -d \tilde t ^2 + \cosh^4 \left(\frac{x}{2} \right) (dx^2 + d \Omega_2^2) \right). \label{conformal_ERN_1}
 \ee
 In contrast to the exterior of the ERN, the spacetime described by the line element in the brackets in equation~\eqref{conformal_ERN_1}, 
 \be
d \tilde{s}^2 = -d \tilde t ^2 + \cosh^4\left(\frac{x}{2} \right)  (dx^2 + d \Omega_2^2), \label{conformal_ERN}
\ee 
is geodesically complete with $(\tilde t , x) \in \mathbb{R}^2$. The Ricci scalar of this spacetime is zero.

This spacetime has two asymptotically flat ends: $x\to \pm \infty$.  Asymptotic flatness at these ends can be seen by introducing, say,  $\rho = \cosh^2 \left(\frac{x}{2} \right) $ and taking $\rho \to \infty$ limit. The reflection symmetry $x \leftrightarrow -x $ is an isometry of the metric  $d \tilde{s}^2 $. The $x \leftrightarrow -x $  is a conformal symmetry of the metric $ds^2$. This symmetry is precisely the CT symmetry discussed in the previous sections.

\begin{figure}[t]
\begin{center}
 \includegraphics[width=0.45\textwidth]{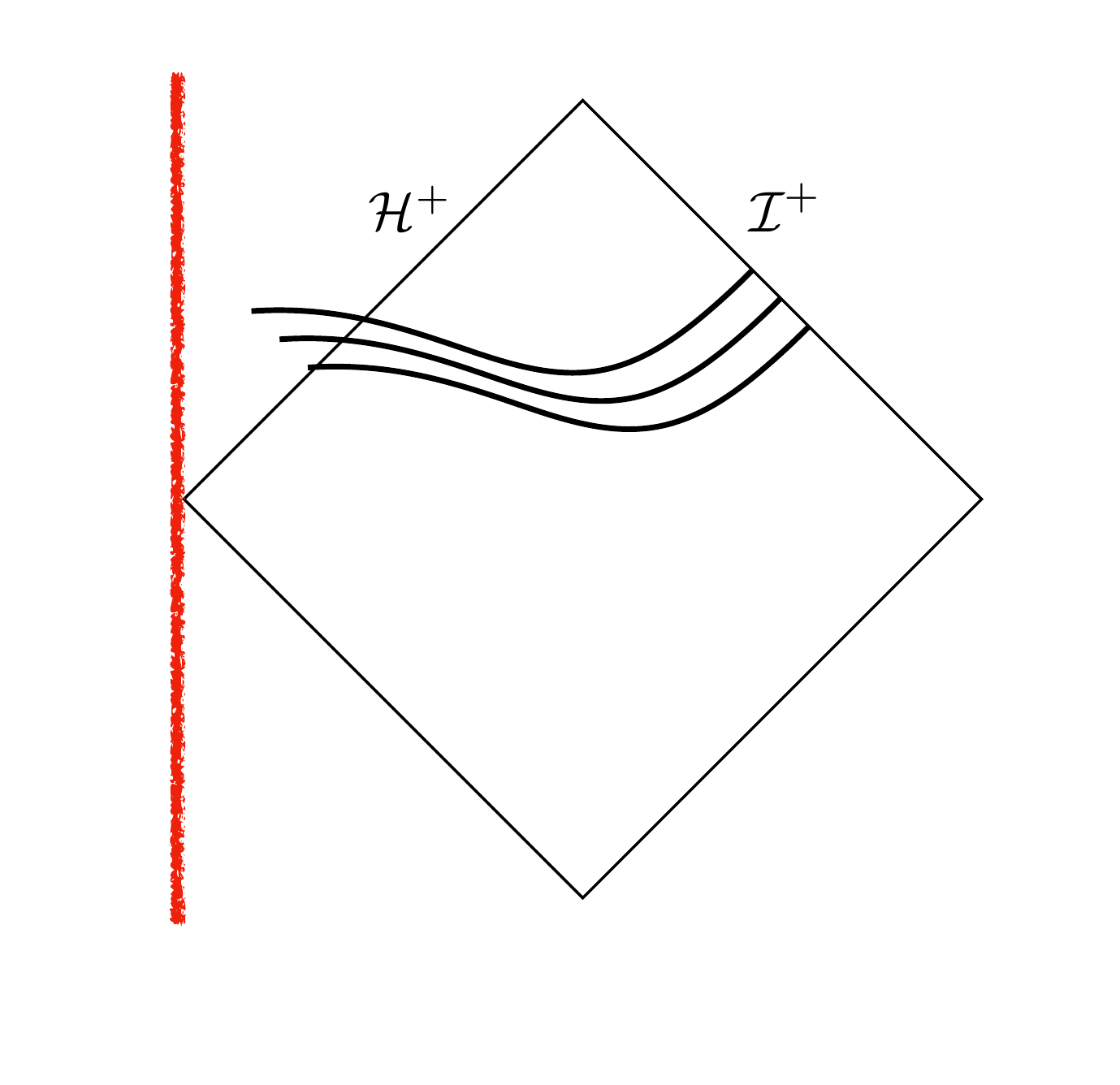}
 \caption{\sf Schematic drawing of constant $s$ spacelike surfaces in the  ERN background. The surfaces reach  future null infinity along $u=s$ outgoing null lines. At the future horizon the surfaces intersect at advanced time $v=s$, however, the normal there is timelike.}
\label{Fig:HyperbolicSlicing}
\end{center}
\end{figure}

Next we introduce,
\be
s = \tilde t   - \frac{1}{2} \left( \cosh x + \ln (2 \cosh x) \right),
\ee
and foliate the spacetime by hyperboloidal surfaces of constant $s$. These surfaces are called hyperboloidal as their asymptotic behavior is similar to standard hyperboloids in Minkowski spacetime~\cite{Zenginoglu:2007jw}.  These surfaces are spacelike. The ERN metric in coordinates 
$(s, x, \theta, \phi)$ takes the form, 
\begin{align}
g_{ss} &= - g , &
g_{xs} &=  - \frac{1}{2} g (\sinh x  +  \tanh x),  & \\
g_{xx} &= g \cosh ^4\left(\frac{x}{2}\right) \text{sech}^2(x),&
g_{z \bar z} &=  \frac{2g}{(1+ z \bar z)^2}.
\end{align}
where we denote the conformal factor as,
\be
g(x) =  \frac{16 M^2}{(1+ e^{-x})^2}.
\ee
The inverse metric components take the form
\begin{align}
g^{ss} &= -g^{-1} \text{sech}^2(x),  &
g^{xs} &= - 2 g^{-1} \text{sech}\  x \tanh \left(\frac{x}{2} \right),  & \\ 
g^{xx} &= g^{-1}  \text{sech}^{4} \left(\frac{x}{2}\right)&
g^{z \bar z} &=  \frac{1}{2} g^{-1} (1+ z \bar z)^2.
\end{align}

\begin{figure}[t]
\begin{center}
 \includegraphics[width=0.45\textwidth]{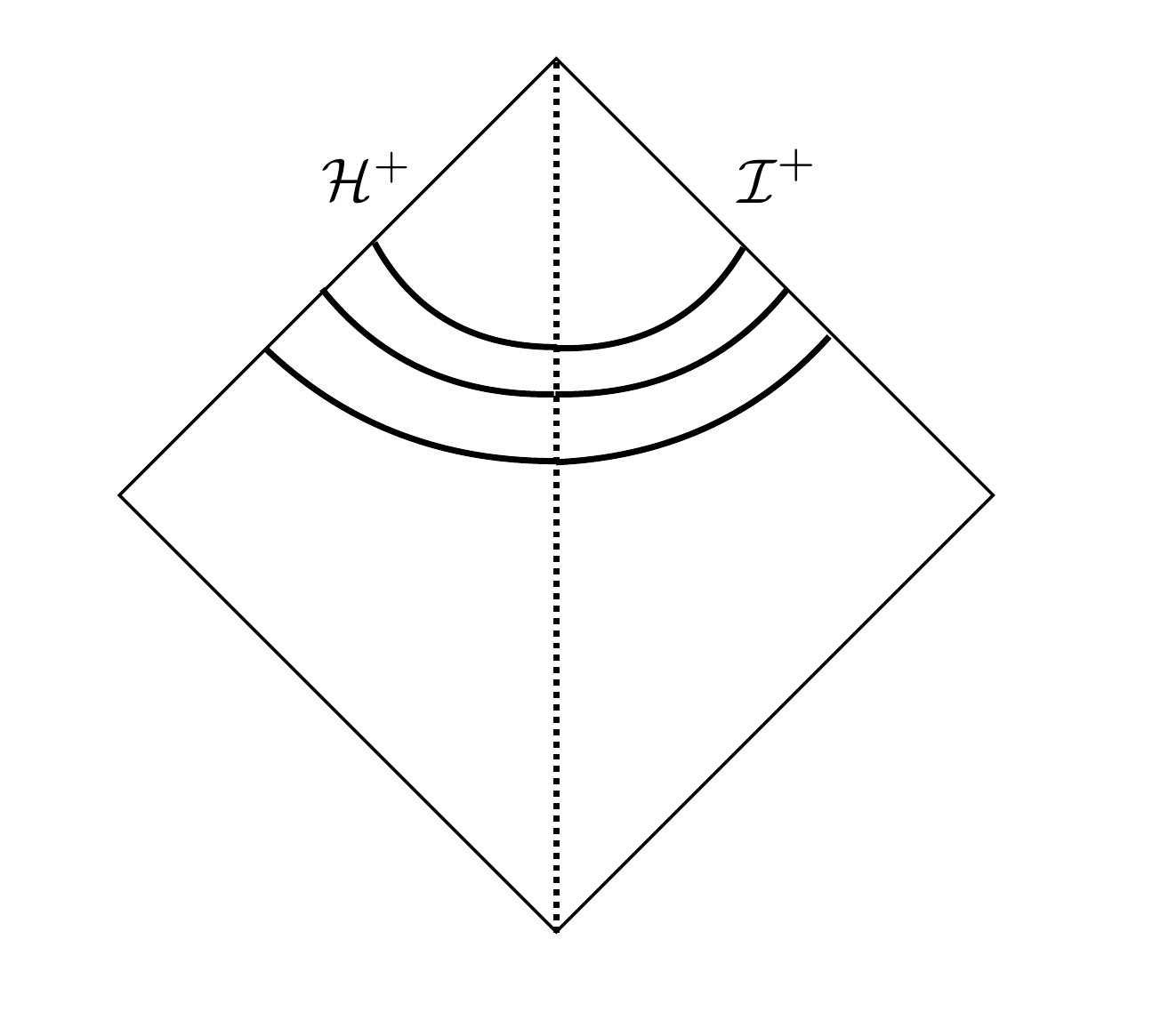}
 \caption{\sf Schematic drawing of constant $s$ spacelike surfaces in the  conformally rescaled ERN background described by the line element \eqref{conformal_ERN}. A constant $s$ surface reaches the right future null infinity along the outgoing null line $u=s$ and the left future null infinity along the ingoing null line $v=s$. }
\label{Fig:HyperbolicSlicing2}
\end{center}
\end{figure}

From the component $g^{ss}$ it follows that the normal to the constant $s$ surface $n_a$ has the norm,
\be
n \cdot n = g^{ss} = -\frac{1}{4 M^2 }\frac{\left(1+ e^x\right)^2}{\left(1+ e^{2 x}\right)^2}.  
\ee
The norm goes to zero as $x\to \infty$. Let us calculate the value of $u$ where $s=\mbox{const}$ hypersurface intersects future null infinity as $x\to \infty$: 
\bea
u &=& t - r_* = 4 M \tau - \left(r-M - \frac{M^2}{r-M} + 2 M \ln \left (\frac{r}{M}-1\right) \right), \\
&=& 2 M \left(2 s + e^{-x} +  \ln \left[ \frac{2 \cosh x}{e^x} \right] \right) .
\eea
Thus,
\be
u \rightarrow  4 M s \qquad  \mbox{as} \qquad  x\to \infty.
\ee
 The $s=\mbox{const}$ hypersurface approaches a finite point at null infinity as $x\to \infty$.  A similar calculation shows, 
\be
v = t + r_*  = 2 M \left( 2 s + e^{x} + \ln \left[2 \cosh x \, e^x \right] \right). 
\ee
Thus, 
\be
v \rightarrow  4 M s \qquad  \mbox{as} \qquad  x\to - \infty.
\ee
i.e., along the $s=\mbox{const}$ hypersurface we approach a finite point at the future horizon $x\to - \infty$. The norm of the normal $n \cdot n $ approaches $-1/(4M^2) $ as  $x\to - \infty$. The slices are schematically shown in figure \ref{Fig:HyperbolicSlicing}. In the unphysical spacetime $d \tilde{s}^2$ the slices become symmetrical with respect to the left and right null infinities. This is shown in figure \ref{Fig:HyperbolicSlicing2}.

We consider the toy model problem (the residual gauge transformations for the harmonic gauge),
\be
\widetilde \Box \epsilon = 0, \label{tilde-box}
\ee 
on the spacetime conformally related to the ERN spacetime described by metric \eqref{conformal_ERN}. 
Introducing $\tilde r \in (- \infty, \infty)$ such that, 
\be
\tilde r = \int \cosh^2\left(\frac{x}{2} \right) dx = \frac{x}{2}+\frac{\sinh (x)}{2},
\ee
and then $\rho \in (- \infty, \infty)$ such that, 
\bea
\tilde t &=& \tau \sqrt{(1+\rho^2)}, \\
\tilde r &=& \rho \tau,
\eea
metric \eqref{conformal_ERN} takes the form, 
\be
d \tilde{s}^2 = -d \tau^2 + \tau^2 \left( \frac{d\rho^2 }{1+\rho^2} + \tau^{-2}  f(\rho, \tau)^2 d \Omega_2^2\right).
\ee
The function $f(\rho, \tau)$ is an implicit function of coordinates $\rho, \tau$,
\be
f(\rho, \tau) = \cosh^2\left(\frac{x (\rho, \tau)}{2} \right). \label{func_f}
\ee

\begin{figure}[t]
\centering
\begin{center}
\includegraphics[width=0.5\textwidth]{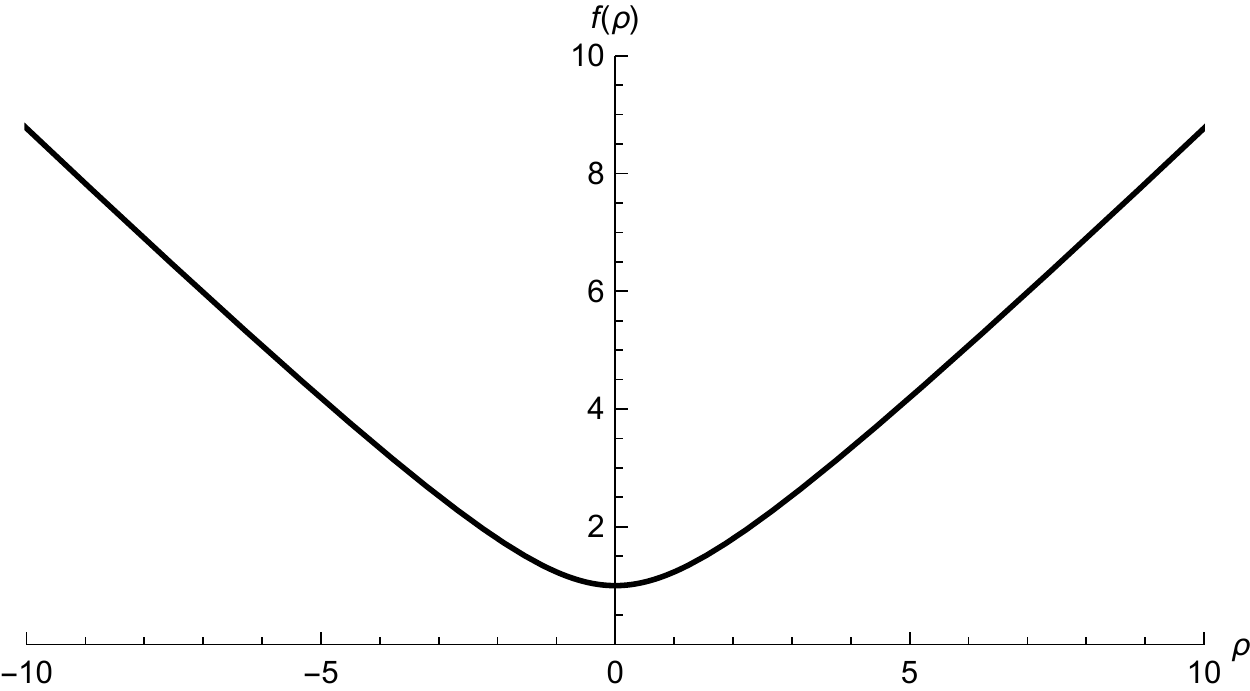}
\caption{\sf  A graph of $f(\rho, \tau)$ as a function of $\rho$ on the $\tau = 1$ slice.}
\label{fig:graph}
\end{center}
\end{figure}

Consider the constant $\tau$ slice, say, $\tau =1$. The metric induced on the slice is 
\be
ds^2_\rom{slice} = \frac{d\rho^2 }{1+\rho^2} +    f(\rho, \tau = 1)^2 d \Omega_2^2.
\ee
A graph of $f(\rho, \tau)$ as a function of $\rho$ on the $\tau = 1$ slice is shown in figure \ref{fig:graph}.  In the limit $\rho \to \pm \infty$,  function $f(\rho, \tau=1)$ behaves as $\lim_{\rho \to \pm \infty} f(\rho, \tau=1) \to |\rho|$. Thus, in the $\rho \to \pm \infty$ limit    the slice is asymptotically Euclidean AdS$_3$. The slice has two asymptotically AdS$_3$ ends.\footnote{Additional motivation for working with such a slicing and harmonic gauge fixing comes from the success of such an approach at  timelike infinity in flat space \cite{Campiglia:2015qka}. Working in the harmonic gauge, Campiglia and Laddha showed that the angle-dependent large gauge transformations introduced at future null infinity have a natural extension in the interior. The slices then can be pushed to timelike infinity.  The gauge transformations at timelike infinity have a well defined action on the asymptotic phase space of massive particles, and the resulting Ward identities are found to be equivalent to Weinberg’s soft photon theorem.}  On the $\tau = 1$ slice consider 
\be
\epsilon(\rho, \theta, \phi) = \sum_{l = 0}^{\infty} \epsilon_l(\rho) Y_{lm}.
\ee
The resulting equation for the function $\epsilon_l(\rho)$ for a fixed $l$ is,
\be
      (1+ \rho ^2 ) f(\rho)^2 \epsilon_l''(\rho ) + f(\rho) ((2 (1+\rho^2) f'(\rho )+\rho f(\rho))) \epsilon_l
   '(\rho )
   -l (l+1) \epsilon_l
   (\rho )=0.
\ee
In the asymptotic region $\rho \to \infty$, the ``non-normalisable'' solution goes as, 
\be
\epsilon_l (\rho) \sim \rho ^l \, _2F_1\left(1+ \frac{1}{2}l,\frac{1}{2}l;l+\frac{3}{2};-\rho ^2\right),
\ee
which becomes a constant in the $\rho \to \infty$ limit. With a numerical integration, it can be readily seen that as $\rho$ goes from negative to positive values a symmetric $\epsilon_l(\rho)$ can be found to smoothly interpolate between the same constant values as $\rho \to \pm \infty$. Thus, by summing over spherical harmonics it follows that for this toy model problem, the gauge parameter functions $\epsilon(z,\bar{z})$ at the two null infinities can be taken to be identical. We conjecture that the same can be done for the Eastwood-Singer gauge condition at the future horizon and future null infinity in the ERN spacetime.

Two comments are in order here.  First, we have presented the above discussion on the $\tau = 1$ slice, but in fact the discussion is independent of any fixed value of $\tau$. To see this note that for any fixed $\tau$, the induced metric on the constant $\tau$ slice is
 \be
 ds^2_\tau = \tau^2 \left( \frac{d\rho^2 }{1+\rho^2} + \tau^{-2}  f(\rho, \tau)^2 d \Omega_2^2\right), \label{metric_constan_tau}
 \ee
where the function $f(\rho, \tau)$ is given in equation \eqref{func_f}. Apart from an overall scaling, the relevant property to understand    is the behavior of $\tau^{-1} f(\rho, \tau)$   as a function of $\rho$. In terms of the variable  $x$ it is clear that 
\be
\left\{\rho, \tau^{-1} f(\rho, \tau)\right\} = \left\{ \frac{1}{\tau} \left( \frac{x}{2}+\frac{\sinh (x)}{2} \right),\frac{1}{\tau}\cosh^2\left(\frac{x }{2}\right)\right\}.
\ee
Figure \ref{fig:graph} is precisely the parametric plot for $\tau =1$ of $\left\{ \frac{1}{\tau} \left( \frac{x}{2}+\frac{\sinh (x)}{2} \right),\frac{1}{\tau}\cosh^2\left(\frac{x }{2}\right)\right\}$. For $\tau \neq 1$ both the $x$ and $y$ axes of the graph get rescaled by a factor of $\frac{1}{\tau}$. In particular, for any $\tau$ in the limit $\rho \to \pm \infty$,  $\tau^{-1} f(\rho, \tau)$ goes as $\rho$. Thus,  $\tau=$ constant slices all have two asymptotically AdS$_3$ ends.

 Second, in the $\tau \to \infty$ limit the metric
 \be
ds^2_{i^+} =   \frac{d\rho^2 }{1+\rho^2} + \tau^{-2}  f(\rho, \tau)^2 d \Omega_2^2,
 \ee
 can be thought of as the blow up of the point $i^+$ for the spacetime described by metric \eqref{conformal_ERN}.\footnote{The blow-up in the sense of Ashtekar-Hansen \cite{Ashtekar:1978zz}. For a recent discussion of this blow-up  in the context of BMS-supertranslations at spacelike infinity see \cite{Prabhu:2019fsp}. } It can be readily checked that for any finite $\rho \neq 0$,  $\tau^{-1}  f(\rho, \tau)$ behaves as $\rho$ in the $\tau \to \infty$ limit. In the neighbourhood of $\rho = 0$ it smoothly interpolates between the $\rho > 0$ and the $\rho < 0$ AdS$_3$ regions.

Studies on interpolating asymptotic dynamics  between two different asymptotic   regions include \cite{Grumiller:2019ygj, Henneaux:2019sjx}. The toy model example studied in this section calls for a corresponding study in four-dimensional asymptotically flat settings. 

\section{Horizon hair and soft charges}
\label{sec:horizon_hair}

For the Maxwell field on flat spacetime, it is now well appreciated that there exists large gauge transformations parametrised by arbitrary non-vanishing gauge parameters on the asymptotic sphere at null infinity \cite{He:2014cra, 1703.05448, Campiglia:2016hvg}.  As a consequence, there exist an infinite number of boundary symmetry charges \cite{Iyer:1994ys, Chandrasekaran:2018aop}. In this section we present an expression for the global charge (often called the Iyer-Wald charge) for the probe Maxwell field on the ERN spacetime.

 For the electromagnetic field, with Lagrangian,
\be
L =- \frac{1}{4} \int \sqrt{-g}  \cF_{ab} \cF^{ab}d^4x,
\ee
the symplectic form as an integral over an arbitrary Cauchy surface $\Sigma$ is,
\begin{equation}
\Omega\left(\mathcal{A}\,,\delta_1\mathcal{A}\,,\delta_2 \mathcal{A} \right) = - \int _{\Sigma}  \left( \delta_1 A^{b} \delta_2 \mathcal{F}_{a b}   - \delta_2 \mathcal{A}^{b} \delta_1 \mathcal{F}_{a b}\right) n^a   \sqrt{h} d^3 x,
\label{cs.symp}
\end{equation}
where $n^{a}$  is the future pointing unit normal to $\Sigma$.  The total charge as the generator of gauge transformations is given as, 
\begin{equation}
\delta Q_{\epsilon} = \Omega\left(\mathcal{A}\,,\delta\mathcal{A}\,,\delta_{\epsilon} \mathcal{A} \right).
\label{cs.dcharge}
\end{equation}
One finds,
\begin{equation}
Q_{\epsilon} =  \int _{\Sigma} \nabla^{b}\left(  \epsilon \mathcal{F}_{ab}\right) n^a   \sqrt{h} d^3 x.
\label{cs.charge}
\end{equation}

On black hole spacetimes and in the absence of massive particles, a choice of Cauchy surface is $\mathcal{I}^+ \cup \mathcal{H}^+$. The Cauchy surface $\mathcal{I}^+ \cup \mathcal{H}^+$ can be reached by taking the $t \to \infty$ limit of constant $t$ Cauchy surfaces $\Sigma_t$ for  the exterior of the ERN black hole.  In previous sections,  we argued that $\epsilon_{\mathcal{H}^+}(z, \bar z) = \epsilon_{\mathcal{I}^+}(z\,\bar z)$: the gauge parameter is the same function of the sphere coordinates at $\mathcal{I}^+$ and at $\mathcal{H}^+$. As a result, the charge integral \eqref{cs.charge} is well defined when we push the Cauchy surface to $\mathcal{I}^+ \cup \mathcal{H}^+$.  The total integral splits into two parts, 
\bea
Q_{\epsilon} &:=& \lim_{\{ \Sigma_t \to \mathcal{I}^+ \cup \mathcal{H}^+ \}} \int _{\Sigma}  \nabla^{b}\left(  \epsilon \mathcal{F}_{ab}\right) n^a   \sqrt{h} d^3 x\\ 
&=& \int _{ \mathcal{I}^+}  \nabla^{b}\left(  \epsilon \mathcal{F}_{ab}\right) n^a   \sqrt{h} d^3 x + \int _{\mathcal{H}^+}   \nabla^{b}\left(  \epsilon \mathcal{F}_{ab}\right) n^a   \sqrt{h} d^3 x \\
& =& Q^{\mathcal{I}^+}_{\epsilon} + Q^{\mathcal{H}^+}_{\epsilon}.
\label{cs.scharge}
\eea

In outgoing Eddington-Finkelstein coordinates \eqref{es.oef},  
$
n^a   \sqrt{h} d^3x = \delta^a_u r^2  \gamma_{z \bar{z}} du dz d \bar{z}.  
$ 
With boundary conditions, 
\bea
  \mathcal{F}_{ij}\left(u\,,r\,,z\,,\bar{z}\right) &=& F^{(0)}_{ij}\left(u\,, z\,, \bar{z}\right)  + \mathcal{O}\left(\frac{1}{r}\right), \\ 
 \mathcal{F}_{i r}\left(u\,,r\,,z\,,\bar{z}\right) &=& \frac{1}{r^2} F^{(0)}_{i r}\left(u\,, z\,, \bar{z}\right)  + \mathcal{O}\left(\frac{1}{r^3}\right),
\label{cs.maxinf}
\eea
where $i, j$ collectively stand for $(u,z,\bar{z})$, the  integrand at $\mathcal{I}^+ $  becomes 
\be
\nabla^{b}\left(  \epsilon \mathcal{F}_{ab}\right) n^a   \sqrt{h} = 
 -\gamma_{z \bar{z}}\,  \epsilon_0\left(z\,,\bar{z}\right) \partial_{u} F^{(0)}_{r u} + \partial_z\left(\epsilon_0\left(z\,,\bar{z}\right) F^{(0)}_{u \bar{z}}\right) + \partial_{\bar{z}}\left(\epsilon_0\left(z\,,\bar{z}\right) F^{(0)}_{u z}\right). 
\ee
Hence the $Q^{\mathcal{I}^+}_{\epsilon}$ contribution in equation~(\ref{cs.scharge}) takes the form, 
\begin{equation}
Q^{\mathcal{I}^+}_{\epsilon} =  - \int _{\mathcal{I}^+} du dz d\bar{z} \,  \gamma_{z \bar{z}}\,  \epsilon_0\left(z\,,\bar{z}\right) \partial_{u} F^{(0)}_{r u}\left(u\,,z\,,\bar{z}\right)\,,
\label{cs.soft}
\end{equation}
where $ \epsilon_0\left(z\,,\bar{z}\right) $ is the limiting value of $\epsilon$ on approaching ${\mathcal{I}^+}$. Equation~(\ref{cs.soft}) is the known expression for the soft charge on asymptotically flat spacetimes in terms of a volume integral over $\mathcal{I}^+$~\cite{He:2014cra, 1703.05448}. 

In ingoing Eddington-Finkelstein coordinates, the integrand of equation~(\ref{cs.charge}) at $\mathcal{H}^+ $ ($r=M$), becomes 
$
n^a   \sqrt{h} d^3x = \delta^a_v M^2  \gamma_{z \bar{z}} dv dz d \bar{z}.  
$ 
With the boundary conditions, 
\bea
  \mathcal{F}_{ij}\left(v\,,r\,,z\,,\bar{z}\right) &=&  \bar {F}^{(0)}_{ij}\left(v\,, z\,, \bar{z}\right)  +  \mathcal{O}(r-M), \\ 
 \mathcal{F}_{i r}\left(v\,,r\,,z\,,\bar{z}\right) &=& \frac{1}{M^2}  \bar {F}^{(0)}_{i r}\left(v\,, z\,, \bar{z}\right)  + \mathcal{O}(r-M) ,
 \label{cs.maxhor}
\eea
where $i, j$ now collectively stand for $(v,z,\bar{z})$, the  integrand at $\mathcal{H}^+ $  becomes 
\be
\nabla^{b}\left(  \epsilon \mathcal{F}_{ab}\right) n^a   \sqrt{h} = 
 \gamma_{z \bar{z}}\,  \epsilon_0\left(z\,,\bar{z}\right) \partial_{v} \bar {F}^{(0)}_{r v} + \partial_z\left(\epsilon_0\left(z\,,\bar{z}\right) \bar {F}^{(0)}_{v \bar{z}}\right) + \partial_{\bar{z}}\left(\epsilon_0\left(z\,,\bar{z}\right) \bar {F}^{(0)}_{v z}\right). 
\ee
 Hence,  $Q^{\mathcal{H}^+}_{\epsilon}$ contribution to equation~(\ref{cs.scharge}) takes the form, 
\begin{equation}
Q^{\mathcal{H}^+}_{\epsilon} =  \int _{\mathcal{H}^+} dv dz d\bar{z} \,  \gamma_{z \bar{z}}\,  \epsilon_0 \left(z\,,\bar{z}\right) \partial_{v} \bar {F}^{(0)}_{r v}\left(v\,,z\,,\bar{z}\right)\,,
\label{cs.hair}
\end{equation}
in terms of a volume integral over $\mathcal{H}^+$.
 
Now we can see that the horizon soft charges follow from the CT dual of the null infinity soft charges. We first note from equation~(\ref{cs.maxinf}) that the CT transformation of $r^2 \mathcal{F}_{r u}\left(u\,,r\,,z\,,\bar{z}\right)$ is, 
\be
\mathcal{T}_*\left(r^2 \mathcal{F}_{r u}\left(u\,,r\,,z\,,\bar{z}\right)\right) = \mathcal{T}_*\left(  F^{(0)} \left(u\,, z\,, \bar{z}\right)  + \mathcal{O}\left(\frac{1}{r}\right)\right) 
\label{ct.fur}
\ee
At leading order in $(r-M)$, this becomes, via equation \eqref{fur.ct},
\be
M^2 \mathcal{F}_{r v}\left(v\,,r\,,z\,,\bar{z}\right)  =  -     F^{(0)}\left(v\,, z\,, \bar{z}\right) + \mathcal{O} \left(\frac{r -M}{M}\right).
\ee
Note that after the transformation, the function $ F^{(0)}$ has arguments $\left(v\,, z\,, \bar{z}\right)$, but  it is otherwise the same function. As a result, 
\bea
Q^{\mathcal{H}^+}_{\epsilon} 
&=& - \int_{\mathcal{H}^+} dv dz d\bar{z} \,  \gamma_{z \bar{z}}\,  \epsilon_0\left(z\,,\bar{z}\right) \partial_{v}  F^{(0)}\left(v\,,z\,,\bar{z}\right) \\
&=& \mathcal{T}_* \left[ \int_{\mathcal{I}^+} du dz d\bar{z} \,  \gamma_{z \bar{z}}\,  \epsilon_0\left(z\,,\bar{z}\right) \partial_{u}  F^{(0)}\left(u\,,z\,,\bar{z}\right)\right] ~=~ \mathcal{T}_* \left(Q^{\mathcal{I}^+}_{\epsilon}\right).
\label{cs.ctsoft}
\eea
Hence soft electric hair on the horizon of the ERN spacetime follow from the CT transformation on the soft electric hair at null infinity.  This is schematically shown in figure \ref{Fig:CT_mapping}.

 It is natural to expect that the conservation law in the present setting takes the form 
\be
Q^{\mathcal{I}^+}_{\epsilon} + Q^{\mathcal{H}^+}_{\epsilon} = Q^{\mathcal{I}^-}_{\epsilon} + Q^{\mathcal{H}^-}_{\epsilon}.
\ee
If not on the ERN spacetime, it should be possible to make precise the conservation law following \cite{Campiglia:2017mua} on the  spacetime with two asymptotic flat ends considered in section \ref{sec:toy_model}.

 \section{Aretakis and Newman-Penrose constants}
 \label{sec:NP_charges}
In this section, we write expressions for  Aretakis and Newman-Penrose constants for a probe Maxwell field in an ERN background and relate them via the inversion symmetry. This discussion is an extension of the scalar analysis of refs.~\cite{Bizon:2012we, Lucietti:2012xr, Sela:2015vua, Godazgar:2017igz, Bhattacharjee:2018pqb}; and is largely independent of the discussion of the previous sections.

We start with the spherical harmonics decomposition of the Maxwell field $\mathcal{A}_a$ in the outgoing Eddington-Finkelstein coordinates cf.~(\ref{es.oef}). Expanding various components of the Maxwell field in appropriate scalar and vector spherical harmonics we have both even $(-1)^l$ and odd $(-1)^{l+1}$ parity terms, $\mathcal{A}_a dx^a =  \mathcal{A}_a^{\mathrm{odd}} dx^a + \mathcal{A}_a^{\mathrm{even}} dx^a$, as
\begin{align}
\mathcal{A}_a^{\mathrm{odd}} dx^a &=\sum_{lm} \alpha_{lm} \left(\partial_{z}Y^{lm} dz - \partial_{\bar{z}}Y^{lm} d\bar{z} \right),\label{a.odd}\\
{\mathcal{A}}_a^{\mathrm{even}}  dx^a &=\sum_{lm}\left(f_{lm}Y^{lm} du + h_{lm}Y^{lm} dr +  \kappa_{lm} (\partial_{z}Y^{lm} dz + \partial_{\bar{z}}Y^{lm} d \bar{z}) \right), \label{a.even}
\end{align}
where $Y_{lm}$ are the standard scalar spherical harmonics satisfying,
\begin{equation}
2 \gamma^{z \bar{z}} \partial_z \partial_{\bar{z}} Y^{lm} = -l (l+1) Y^{lm}.
\end{equation}
The coefficients in the decomposition, $\alpha_{lm}\,,f_{lm}\,,h_{lm}$ and $\kappa_{lm}$ are functions of $r,u$. For now we restrict ourselves to $l\ge 1$,  we comment on the $l=0$ case separately below.  Even and odd parity perturbations can be fully described by one gauge invariant variable each. These variables satisfy a decoupled wave equation~\cite{Ruffini:1972pw} of the form,
\begin{equation}
2 \partial_u\partial_r \psi_{l} - \partial_r \left(g^{rr} \partial_r \psi_{l}\right) + \frac{l(l+1)}{r^2}\psi_{l} = 0\, , \label{max.wav1}
\end{equation}
 in outgoing Eddington-Finkelstein coordinates,
and of the form, 
\begin{equation}
2 \partial_v\partial_r \psi_{l} + \partial_r \left(g^{rr} \partial_r \psi_{l}\right) - \frac{l(l+1)}{r^2}\psi_{l} = 0 \, , \label{max.wav2} 
\end{equation}
in ingoing Eddington-Finkelstein coordinates.
For the odd  parity perturbation $\psi_{l} = \alpha_{lm}$ and for the even parity perturbation $\psi_{l} = \frac{1}{l(l+1)} r^2  (\partial_u h_{lm} - \partial_r f_{lm})$.  A small calculation shows that,
\bea
{\cal{F}}_{z \bar{z}}^{\text{odd}} &=&- 2\alpha_{lm}\partial_z \partial_{\bar{z}}Y^{lm} \label{max.odd1}, \\
{\cal{F}}_{ur}^{\text{even}} &=& \left(\partial_u h_{lm}-\partial_r f_{lm}\right)Y^{lm} \label{max.even1}.
\eea
Thus the $\psi_l$ entering in the above equations are essentially (upto numerical factors) the magnetic component ${\cal{F}}_{z \bar{z}}$ for the odd parity field and $r^2$ times the electric field component ${\cal{F}}_{ur}$ for the even parity field. Let us denote the even and odd parity fields as $\psi_l^+$ and $\psi_l^-$. In situations where the distinction is not relevant, we simply denote the two fields collectively as $\psi_l$.

We first demonstrate that the wave equation \eqref{max.wav2} in ingoing Eddington-Finkelstein coordinates admits an infinite tower of Aretakis constants, one for each $l$ at  $\mathcal{H}^+$. Our construction parallels the corresponding discussion in section 6.2 of \cite{Lucietti:2012xr}. 
Let $f(r)$ be a smooth function that is non-vanishing at the horizon, and without loss of generality we set the function $f(r)\big{|}_{r=M} = 1.$ Multiplying equation \eqref{max.wav2} by $r^2 f(r)$ and differentiating $l$ times with respect to $r$ and evaluating at $r=M$ we deduce that,
\be
A_l [\psi_l] = \frac{M^{l-1}}{(l+1)!}[\partial_r^l (r^2 f(r) \partial_r \psi)] \bigg{|}_{r=M},
\label{art.consts}
\ee
is conserved along $\cal{H}^+$ for $l > 0$, provided the derivatives of the function $f(r)$ at $r=M$ are related by the following set of equations,
\begin{equation}
f^{(k)}\Big \vert_{r=M} = - \frac{2 (l-k)}{2l + 1 - k} \left(r^{-1} f \right)^{k-1}\Big \vert_{r=M},
\label{art.cond}
\end{equation}
for $1 \le k \le l$. Equations  \eqref{art.cond} determine the constants $f^{(k)}\Big \vert_{r=M}$ recursively. The Aretakis constants 
\eqref{art.consts}
only depend on these derivatives and are independent of the specific choice of the function $f(r)$.  A rich class of configurations can be described by an expansion in powers of $(r-M)$ near the horizon as,  
\begin{align}
\psi_{l}(v,r) &= \sum\limits_{k=0}^{\infty} a_{k}(v) \frac{\left(r-M\right)^k}{M^k}. \label{psi.he}
\end{align}
We can readily calculate the form of the Aretakis charges for the solution of the form (\ref{psi.he}). It gives,
\be
A_l = a_l + a_{l+1} \qquad \mbox{for} \qquad l  \ge 1.
\ee

We now construct the Newman-Penrose constants.  A rich class of configurations can be described as an expansion in inverse powers of $r$ near null infinity   in outgoing coordinates as,
\begin{equation}
\psi_{l}(u,r) = \sum\limits_{k=0}^{\infty} b_{k}(u) \left(\frac{M}{r}\right)^{k}.
\label{psi.ie}
\end{equation}
Inserting this expansion into equation \eqref{max.wav1} and looking at successive inverse powers of $r$ gives a set of linear equations. These equations can be expressed concisely in terms of Pascal matrices as first discussed in \cite{Bhattacharjee:2018pqb}. We follow the same strategy. For a given $l$ we look at the set of equations in powers of $r$ coming from the first $l+2$ terms in the expansion \eqref{psi.ie}, i.e., the first $l+1$ equations involving $b_0, b_1, \ldots, b_l, b_{l+1}$. We organise these equations using $(l+1) \times (l+1)$ matrices whose components are labelled by $i=0,1, \ldots, l$. We consider the vector $\mathbf{b}$  whose components are $b_i$, and the vector $\dot{\mathbf{b}}_+$ whose components are $(\dot{\mathbf{b}}_+)_i = \partial_u b_{i+1}$. The equations of motion can then be summarised as
\be
M \mathsf{N}_l \dot{\mathbf{b}}_+ = [\tfrac{1}{2} l (l + 1) - \mathsf{P}_l] \mathbf{b} ,
\ee
where $ \mathsf{N}_l$ is the diagonal matrix $(\mathsf{N}_l)_{ij} = (i+1) \delta_{ij}$ and $\mathsf{P}_l$ is a lower triangular matrix with entries $(\mathsf{P}_l)_{ij} = \frac{1}{2} i (i+1) \delta_{i,j} - (i^2-1) \delta_{i, j+1}  + \frac{1}{2} (i-2)(i+1) \delta_{i,j+2}$.

The matrix $\mathsf{P}_l$ can be diagonalised as
\begin{equation}
\mathsf{P}_l = \mathsf{L}_l \mathsf{T}_l \mathsf{L}_l^{-1}\,,
\label{np.sim}
\end{equation}
where $\mathsf{T}_l$ is a diagonal matrix with entries $(\mathsf{T}_l)_{ij} = \frac{1}{2}i(i+1) \delta_{i,j}$. The matrix $\mathsf{L}_l$ can be written as a product of two matrices $\mathsf{J}_l$ and $\widetilde{\mathsf{L}}_l$, $\mathsf{L}_l = \mathsf{J}_l \widetilde{\mathsf{L}}_l$.  The matrix $\mathsf{J}_l$ has components 
\begin{equation}
(\mathsf{J}_l)_{ij} = (i+1) \delta_{i,j}\,,
\label{np.j}
\end{equation}
 while $\widetilde{\mathsf{L}}_l$ has components,
 \be
\widetilde{\mathsf{L}}_l = \begin{pmatrix}
1 & 0 \\
0 & {} ^{i-1}C_{j-1}
\end{pmatrix} , \quad \text{for} \quad  1\le i,j \le l, 
\ee
where $^{p}C_{q}$ are the binomial coefficients. 
 The $(\widetilde{\mathsf{L}}_l)_{ij}$ components for $i,j \ge 1$ are those of the Pascal matrices~\cite{edestr}. The inverse of $\mathsf{L}_l$ is thus $\mathsf{L}_l^{-1} = (\widetilde{\mathsf{L}}_l)^{-1} \mathsf{J}_l^{-1}$, where $\mathsf{J}_l^{-1}$ has components $ (\mathsf{J}_l^{-1})_{ij} = \frac{1}{i+1} \delta_{i,j}$, while $(\widetilde{\mathsf{L}}_l)^{-1} $ has components
 \be
\widetilde{\mathsf{L}}_l = \begin{pmatrix}
1 & 0 \\
0 & {}  (-1)^{i+j-2} \cdot {}^{i-1}C_{j-1}
\end{pmatrix} , \quad \text{for} \quad  1\le i,j \le l, 
\ee
 It follows that
\be
M \mathsf{L}_l^{-1} \mathsf{N}_l \dot{\mathbf{b}}_+ = [\tfrac{1}{2} l (l + 1) - \mathsf{T}_l] \mathsf{L}_l^{-1} \mathbf{b}.
\label{dotd}
\ee
Since, the last component of the matrix $\mathsf{T}_l$ is $\tfrac{1}{2} l (l + 1) $, the right hand side of the last component of this matrix equation is zero. It implies conservation of $(\mathsf{L}_l^{-1} \mathsf{N}_l \mathbf{b}_+)_l$. A short calculation shows that this quantity is,
\bea
& & N_l ~=  ~ \sum\limits_{i=1}^{l} (-1)^{l+i-2} \cdot  {}^{l-1}C_{i-1}  b_{i+1}, \quad \mbox{for} \quad l \ge 1, \\
& &  \partial_u N_l ~=~ 0.
 \eea
  The constants $N_l$ at null infinity are called the Newman-Penrose constants. Newman and Penrose in~\cite{Newman:1968uj} wrote their expressions as  surface integrals over $\mathcal{I}^+$ for arbitrary $l$. They can be seen to be  related to the constants derived above.

Let us now comment on the $l=0$ mode.  For $l=0$ we only have the even field component,
\be
\mathcal{A}_a^{\text{even}} = \left(f_{00} (u,r), h_{00} (u,r),0,0\right).
\ee
This provides only one Maxwell field component,
\be
\mathcal{F}_{ur} = \left(\partial_u h_{00} - \partial_r f_{00}\right)  =: \beta_{00} .
\ee
Maxwell's equations simply become,
\begin{align}
\partial_r \left(r^2 \beta_{00} \right) &= 0, \notag\\
\partial_u\left(\beta_{00}\right) &=0.
\end{align}
These equations have the solution
\be
\beta_{00} = \frac{c}{r^2},
\ee
where $c$ is a constant --- the electric charge. A similar analysis holds for the ingoing Eddington-Finkelstein coordinates.  Thus, the $l=0$ Aretakis constant and the $l=0$ Newman-Penrose can be taken to be  the electric charge.

We now discuss how  the Aretakis charges map to Newman-Penrose charges under the CT transformation. To do so, we recall that a solution of the probe Maxwell field near $\mathcal{I}^+$ can be determined from a known solution at $\mathcal{H}^+$ and vice-versa. The CT transformation on $\mathcal{F}_{r u}$ acts as, cf.~\eqref{fur.ct},
\begin{equation}
\mathcal{T}_*\left(r^2 \mathcal{F}_{r u} dr du\right) \to - \mathcal{T}_*\left(r^2 \mathcal{F}_{r v}\right) dr dv,
\end{equation}
and the other non-radial components of the electromagnetic field are unaffected. As a result, the even parity field $\psi_l^+$ picks up an additional minus sign and the odd parity field remains the same. With this understanding, let us now work out the CT transformation of the class of configurations described by equation~(\ref{psi.he}). We have,
\begin{align}
\psi_{l}(u,r) & = \mathcal{T}_*\psi_{l}(v,r) = \mathcal{T}_*\left(\sum\limits_{k=0}^{\infty} a_{k}(v) \left(\frac{r -M}{M}\right)^k\right) \notag\\
& = \sum\limits_{k=0}^{\infty} a_{k}(u) \left(\frac{M}{r}\right)^k\left(1 - \frac{M}{r}\right)^{-k}\notag\\
&=\left[a_0 + a_1\frac{M}{r} + (a_1 + a_2)\left(\frac{M}{r}\right)^2 + (a_1 + 2 a_2 + a_3)\left(\frac{M}{r}\right)^3 + \cdots\right]
\label{map.psi}.
\end{align} 
Comparing equation~(\ref{map.psi}) with equation~(\ref{psi.ie}), we find the transformation between coefficients $a_i$ and $b_i$,
\be
{\bf{b}} = \widetilde{\mathsf{L}}_l {\bf{a}},
\ee
which implies,
\begin{equation}
{\bf{b}}_+=  {\mathsf{L}}_l {\bf{a}}_+,
\label{map.atb}
\end{equation}
where $ {\mathsf{L}}_l$ is simply the lower triangular Pascal matrix. In the discussion after equation~(\ref{dotd}) we noted  that the $(l+1)$-th component of the column matrix $\mathsf{L}_l^{-1} \mathsf{N}_l \dot{\mathbf{b}}_+$  provides the Newman-Penrose constants. It then follows that the Newman-Penrose constants for the transformed solutions are the $(l+1)$-th component of the vector,
\be
\mathsf{L}_l^{-1} \mathsf{N}_l  {\mathsf{L}}_l {\bf{a}}_+.
\ee
A short calculation gives the Newman-Penrose constants for the transformed configuration as
\be
N_l = a_l + a_{l+1},
\ee
which are nothing but the Aretakis constants.

Finally, let us discuss the time-independent solutions of the scalar wave equation. 
We consider the wave equation in $\{t\,,r\,,z\,,\bar{z}\}$ coordinates.  For $l \ge 1$, the odd and even parity equations both take the form,
\begin{align}
\left[-\partial_t^2 + \partial_{r_*}^2\right] \psi_{l}(t,r) &= g^{rr}\frac{l\left(l+1\right)}{r^2} \psi_{l}(t,r).
\label{sm.eq}
\end{align}
The $l \neq 0$ time independent solutions of equation~(\ref{sm.eq}) are 
\begin{align}
\psi_l(r) &= \frac{1}{(r-M)^{l+1}}\left(\left(l+1\right)r - M\right),\\
\psi_l(r) & = (r-M)^l (M+ lr).
\end{align}
Under the CT transformation, one static solution goes to the other,
\begin{equation}
 \mathcal{T}_*\left((r-M)^l (M+ lr)\right) = \frac{M^{2l+1}}{(r-M)^{l+1}}\left(\left(l+1\right)r - M\right).
\end{equation}

\section{Conclusions}

In this paper, we have investigated certain properties of solutions of a probe Maxwell field on the exterior of the extreme Reissner-Nordstr\"om (ERN) black hole spacetime.  We  demonstrated in section \ref{sec:CT} that Maxwell's equations are invariant under the Couch-Torrence (CT) transformation of the spacetime, which maps the future null infinity to the future event horizon and vice versa. This in particular implies that asymptotic solutions of Maxwell's equations at null infinity can be mapped to analogous solutions near the event horizon. 

In section \ref{sec:ES_gauge}, we showed that the Eastwood-Singer conformally invariant gauge fixing~\cite{Eastwood:1985eh} is invariant under the CT symmetry of the spacetime. In Eddington-Finkelstein coordinates, using the known asymptotic solutions of the residual gauge parameters at null infinity~\cite{He:2014cra, 1703.05448}, we demonstrated that solutions of the residual gauge parameters near the future event horizon have the same exact form as the solutions at null infinity. 

This raises the interesting question on whether there exists a smooth interpolation in the bulk between solutions of the residual gauge parameters at the future event horizon and at future null infinity. We argued in the affirmative for the existence of such an interpolation through our analysis in section \ref{sec:toy_model}. In this section, we mostly studied a toy model problem. It will be interesting to further understand these bulk interpolating solutions. More generally, it will be interesting to understand asymptotic symmetries for spacetimes with two (or more) asymptotically flat ends.

We investigated conserved charges for the probe Maxwell field on the ERN spacetime in section~\ref{sec:horizon_hair}. We used the expression for the Iyer-Wald charge to define globally conserved charges for the probe Maxwell field on arbitrary Cauchy slices of the spacetime. One such slice is the union of the future null infinity and the future event horizon.   
We argued that, soft electric charges on the horizon of the ERN spacetime follow from the CT transformation on the soft electric charges at null infinity. Soft electric charges on the event horizon are often called soft horizon hair. This is schematically shown in figure \ref{Fig:CT_mapping}.

Finally, in section \ref{sec:NP_charges} we constructed  Newman-Penrose and Aretakis like constants along future null infinity and the future event horizon, respectively.  We showed that these constants are  related via the CT inversion symmetry.

\subsection*{Acknowledgments} We thank Miguel Campiglia and Alok Laddha  for discussions.  Our work is supported in part by the Max Planck Partnergroup ``Quantum Black Holes'' between CMI Chennai and AEI Potsdam and by a grant to CMI from the Infosys Foundation.


\begin{thebibliography}{99}

\bibitem{Bondi:1962px} 
  H.~Bondi, M.~G.~J.~van der Burg and A.~W.~K.~Metzner,
  ``Gravitational waves in general relativity. 7. Waves from axisymmetric isolated systems,''
  Proc.\ Roy.\ Soc.\ Lond.\ A {\bf 269}, 21 (1962).
  doi:10.1098/rspa.1962.0161


\bibitem{Sachs:1962wk} 
  R.~K.~Sachs,
  ``Gravitational waves in general relativity. 8. Waves in asymptotically flat space-times,''
  Proc.\ Roy.\ Soc.\ Lond.\ A {\bf 270}, 103 (1962).
  doi:10.1098/rspa.1962.0206


\bibitem{Sachs:1962zza} 
  R.~Sachs,
  ``Asymptotic symmetries in gravitational theory,''
  Phys.\ Rev.\  {\bf 128}, 2851 (1962).
  doi:10.1103/PhysRev.128.2851


\bibitem{He:2014cra} 
  T.~He, P.~Mitra, A.~P.~Porfyriadis and A.~Strominger,
  ``New Symmetries of Massless QED,''
  JHEP {\bf 1410}, 112 (2014)
  doi:10.1007/JHEP10(2014)112
  [arXiv:1407.3789 [hep-th]].


\bibitem{Kapec:2015ena} 
  D.~Kapec, M.~Pate and A.~Strominger,
  ``New Symmetries of QED,''
  Adv.\ Theor.\ Math.\ Phys.\  {\bf 21}, 1769 (2017)
  doi:10.4310/ATMP.2017.v21.n7.a7
  [arXiv:1506.02906 [hep-th]].


\bibitem{1703.05448} 
  A.~Strominger,
  ``Lectures on the Infrared Structure of Gravity and Gauge Theory,''
  arXiv:1703.05448 [hep-th].


\bibitem{Koga:2001vq} 
  J.~i.~Koga,
  ``Asymptotic symmetries on Killing horizons,''
  Phys.\ Rev.\ D {\bf 64}, 124012 (2001)
  doi:10.1103/PhysRevD.64.124012
  [gr-qc/0107096].



\bibitem{Donnay:2015abr} 
  L.~Donnay, G.~Giribet, H.~A.~Gonzalez and M.~Pino,
  ``Supertranslations and Superrotations at the Black Hole Horizon,''
  Phys.\ Rev.\ Lett.\  {\bf 116}, no. 9, 091101 (2016)
  doi:10.1103/PhysRevLett.116.091101
  [arXiv:1511.08687 [hep-th]].


\bibitem{Mao:2016pwq} 
  P.~Mao, X.~Wu and H.~Zhang,
 ``Soft hairs on isolated horizon implanted by electromagnetic fields,''
  Class.\ Quant.\ Grav.\  {\bf 34}, no. 5, 055003 (2017)
  doi:10.1088/1361-6382/aa59da
  [arXiv:1606.03226 [hep-th]].


\bibitem{Donnay:2016ejv} 
  L.~Donnay, G.~Giribet, H.~A.~Gonz\'alez and M.~Pino,
  ``Extended Symmetries at the Black Hole Horizon,''
  JHEP {\bf 1609}, 100 (2016)
  doi:10.1007/JHEP09(2016)100
  [arXiv:1607.05703 [hep-th]].


\bibitem{Hawking:2016msc} 
  S.~W.~Hawking, M.~J.~Perry and A.~Strominger,
  ``Soft Hair on Black Holes,''
  Phys.\ Rev.\ Lett.\  {\bf 116}, no. 23, 231301 (2016)
  doi:10.1103/PhysRevLett.116.231301
  [arXiv:1601.00921 [hep-th]].


\bibitem{Hawking:2016sgy} 
  S.~W.~Hawking, M.~J.~Perry and A.~Strominger,
  ``Superrotation Charge and Supertranslation Hair on Black Holes,''
  JHEP {\bf 1705}, 161 (2017)
  doi:10.1007/JHEP05(2017)161
  [arXiv:1611.09175 [hep-th]].


\bibitem{Carlip:2017xne} 
  S.~Carlip,
 ``Black Hole Entropy from Bondi-Metzner-Sachs Symmetry at the Horizon,''
  Phys.\ Rev.\ Lett.\  {\bf 120}, no. 10, 101301 (2018)
  doi:10.1103/PhysRevLett.120.101301
  [arXiv:1702.04439 [gr-qc]].


\bibitem{Blau:2015nee} 
  M.~Blau and M.~O'Loughlin,
  ``Horizon Shells and BMS-like Soldering Transformations,''
  JHEP {\bf 1603}, 029 (2016)
  doi:10.1007/JHEP03(2016)029
  [arXiv:1512.02858 [hep-th]].

S.~Bhattacharjee and A.~Bhattacharyya,
``Soldering freedom and Bondi-Metzner-Sachs-like transformations,''
Phys. Rev. D \textbf{98}, no.10, 104009 (2018)
doi:10.1103/PhysRevD.98.104009
[arXiv:1707.01112 [hep-th]].

\bibitem{Penna:2017bdn} 
  R.~F.~Penna,
  ``Near-horizon BMS symmetries as fluid symmetries,''
  JHEP {\bf 1710}, 049 (2017)
  doi:10.1007/JHEP10(2017)049
  [arXiv:1703.07382 [hep-th]].



\bibitem{Grumiller:2018scv} 
  D.~Grumiller and M.~M.~Sheikh-Jabbari,
  ``Membrane Paradigm from Near Horizon Soft Hair,''
  Int.\ J.\ Mod.\ Phys.\ D {\bf 27}, no. 14, 1847006 (2018)
  doi:10.1142/S0218271818470065
  [arXiv:1805.11099 [hep-th]].


\bibitem{Chandrasekaran:2018aop} 
  V.~Chandrasekaran, \'E.~\'E.~Flanagan and K.~Prabhu,
  ``Symmetries and charges of general relativity at null boundaries,''
  JHEP {\bf 1811}, 125 (2018)
  doi:10.1007/JHEP11(2018)125
  [arXiv:1807.11499 [hep-th]].

\bibitem{Donnay:2018ckb}
L.~Donnay, G.~Giribet, H.~A.~Gonz\'alez and A.~Puhm,
``Black hole memory effect,''
Phys. Rev. D \textbf{98}, no.12, 124016 (2018)
doi:10.1103/PhysRevD.98.124016
[arXiv:1809.07266 [hep-th]].


\bibitem{Chen:2020nyh} 
  L.~Q.~Chen, W.~Z.~Chua, S.~Liu, A.~J.~Speranza and B.~d.~S.~L.~Torres,
  ``Virasoro hair and entropy for axisymmetric Killing horizons,''
  arXiv:2006.02430 [hep-th].


\bibitem{Perry:2020ndy} 
  M.~Perry and M.~J.~Rodriguez,
  ``Central Charges for AdS Black Holes,''
  arXiv:2007.03709 [hep-th].


\bibitem{Donnay:2020yxw} 
  L.~Donnay, G.~Giribet and J.~Oliva,
  ``Horizon symmetries and hairy black holes in AdS,''
  arXiv:2007.08422 [hep-th].


\bibitem{Compere:2016hzt} 
  G.~Comp\`ere and J.~Long,
  ``Classical static final state of collapse with supertranslation memory,''
  Class.\ Quant.\ Grav.\  {\bf 33}, no. 19, 195001 (2016)
  doi:10.1088/0264-9381/33/19/195001
  [arXiv:1602.05197 [gr-qc]].


\bibitem{ct:1984}
W. E. Couch and R. J. Torrence, ``Conformal invariance under spatial inversion of extreme
Reissner-Nordstr\"om black holes," Gen. Rel. Grav. 16, 789 (1984).


\bibitem{Blaksley:2007ak} 
  C.~J.~Blaksley and L.~M.~Burko,
  ``The Late-time tails in the Reissner-Nordstr\"om spacetime revisited,''
  Phys.\ Rev.\ D {\bf 76}, 104035 (2007)
  doi:10.1103/PhysRevD.76.104035
  [arXiv:0710.2915 [gr-qc]].


\bibitem{Bizon:2012we} 
  P.~Bizo\'n and H.~Friedrich,
  ``A remark about wave equations on the extreme Reissner-Nordstr\"om black hole exterior,''
  Class.\ Quant.\ Grav.\  {\bf 30}, 065001 (2013)
  doi:10.1088/0264-9381/30/6/065001
  [arXiv:1212.0729 [gr-qc]].
  
  
\bibitem{Lucietti:2012xr} 
  J.~Lucietti, K.~Murata, H.~S.~Reall and N.~Tanahashi,
  ``On the horizon instability of an extreme Reissner-Nordstr\"om black hole,''
  JHEP {\bf 1303}, 035 (2013)
  doi:10.1007/JHEP03(2013)035
  [arXiv:1212.2557 [gr-qc]].


\bibitem{Sela:2015vua} 
   A.~Ori,
  ``Late-time tails in extremal Reissner-Nordstr\"om spacetime,''
  arXiv:1305.1564 [gr-qc].
   O.~Sela,
  ``Late-time decay of perturbations outside extremal charged black hole,''
  Phys.\ Rev.\ D {\bf 93}, no. 2, 024054 (2016)
  doi:10.1103/PhysRevD.93.024054
  [arXiv:1510.06169 [gr-qc]].


\bibitem{Godazgar:2017igz} 
  H.~Godazgar, M.~Godazgar and C.~N.~Pope,
  ``Aretakis Charges and Asymptotic Null Infinity,''
  Phys.\ Rev.\ D {\bf 96}, no. 8, 084055 (2017)
  doi:10.1103/PhysRevD.96.084055
  [arXiv:1707.09804 [hep-th]].


\bibitem{Bhattacharjee:2018pqb} 
  S.~Bhattacharjee, B.~Chakrabarty, D.~D.~K.~Chow, P.~Paul and A.~Virmani,
  ``On late time tails in an extreme Reissner-Nordstr\"om black hole: frequency domain analysis,''
  Class.\ Quant.\ Grav.\  {\bf 35}, no. 20, 205002 (2018)
  doi:10.1088/1361-6382/aade59
  [arXiv:1805.10655 [gr-qc]].


\bibitem{Cvetic:2018gss}
M.~Cveti\v{c} and A.~Satz,
``General relation between Aretakis charge and Newman-Penrose charge,''
Phys. Rev. D \textbf{98}, no.12, 124035 (2018)
doi:10.1103/PhysRevD.98.124035
[arXiv:1811.05627 [hep-th]].



\bibitem{Angelopoulos:2018uwb} 
  Y.~Angelopoulos, S.~Aretakis and D.~Gajic,
  ``Late-time asymptotics for the wave equation on extremal Reissner-Nordstr\"om backgrounds,''
  arXiv:1807.03802 [gr-qc].


\bibitem{Cote:2019kbg} 
  J.~C\^ot\'e, V.~Faraoni and A.~Giusti,
  ``Revisiting the conformal invariance of Maxwell's equations in curved spacetime,''
  Gen.\ Rel.\ Grav.\  {\bf 51}, no. 9, 117 (2019)
  doi:10.1007/s10714-019-2599-x
  [arXiv:1905.09968 [gr-qc]].


\bibitem{Eastwood:1985eh} 
  M.~G.~Eastwood and M.~Singer,
  ``A Conformally Invariant Maxwell Gauge,''
  Phys.\ Lett.\ A {\bf 107}, 73 (1985).
  doi:10.1016/0375-9601(85)90198-7


\bibitem{Bizon:2016flx} 
  P.~Bizoń and M.~Kahl,
  ``A Yang--Mills field on the extremal Reissner-Nordstr\"om black hole,''
  Class.\ Quant.\ Grav.\  {\bf 33}, no. 17, 175013 (2016)
  doi:10.1088/0264-9381/33/17/175013
  [arXiv:1603.04795 [gr-qc]].


\bibitem{Zenginoglu:2007jw} 
  A.~Zengino\u{g}lu,
  ``Hyperboloidal foliations and scri-fixing,''
  Class.\ Quant.\ Grav.\  {\bf 25}, 145002 (2008)
  doi:10.1088/0264-9381/25/14/145002
  [arXiv:0712.4333 [gr-qc]].


\bibitem{Campiglia:2015qka} 
  M.~Campiglia and A.~Laddha,
  ``Asymptotic symmetries of QED and Weinberg's soft photon theorem,''
  JHEP {\bf 1507}, 115 (2015)
  doi:10.1007/JHEP07(2015)115
  [arXiv:1505.05346 [hep-th]].


\bibitem{Ashtekar:1978zz}
A.~Ashtekar and R.~O.~Hansen,
``A unified treatment of null and spatial infinity in general relativity. I - Universal structure, asymptotic symmetries, and conserved quantities at spatial infinity,''
J. Math. Phys. \textbf{19}, 1542-1566 (1978)
doi:10.1063/1.523863

\bibitem{Prabhu:2019fsp}
K.~Prabhu,
``Conservation of asymptotic charges from past to future null infinity: Supermomentum in general relativity,''
JHEP \textbf{03}, 148 (2019)
doi:10.1007/JHEP03(2019)148
[arXiv:1902.08200 [gr-qc]].


\bibitem{Grumiller:2019ygj} 
  D.~Grumiller, M.~M.~Sheikh-Jabbari, C.~Troessaert and R.~Wutte,
  ``Interpolating Between Asymptotic and Near Horizon Symmetries,''
  JHEP {\bf 2003}, 035 (2020)
  doi:10.1007/JHEP03(2020)035
  [arXiv:1911.04503 [hep-th]].


\bibitem{Henneaux:2019sjx} 
  M.~Henneaux, W.~Merbis and A.~Ranjbar,
  ``Asymptotic dynamics of AdS$_3$ gravity with two asymptotic regions,''
  JHEP {\bf 2003}, 064 (2020)
  doi:10.1007/JHEP03(2020)064
  [arXiv:1912.09465 [hep-th]].


\bibitem{Campiglia:2016hvg} 
  M.~Campiglia and A.~Laddha,
  ``Subleading soft photons and large gauge transformations,''
  JHEP {\bf 1611}, 012 (2016)
  doi:10.1007/JHEP11(2016)012
  [arXiv:1605.09677 [hep-th]].


\bibitem{Iyer:1994ys} 
  V.~Iyer and R.~M.~Wald,
  ``Some properties of Noether charge and a proposal for dynamical black hole entropy,''
  Phys.\ Rev.\ D {\bf 50}, 846 (1994)
  doi:10.1103/PhysRevD.50.846
  [gr-qc/9403028].


\bibitem{Campiglia:2017mua}
M.~Campiglia and R.~Eyheralde,
``Asymptotic U(1) charges at spatial infinity,''
JHEP \textbf{11}, 168 (2017)
doi:10.1007/JHEP11(2017)168
[arXiv:1703.07884 [hep-th]].


 
\bibitem{Ruffini:1972pw} 
  R.~Ruffini, J.~Tiomno and C.~V.~Vishveshwara,
  ``Electromagnetic field of a particle moving in a spherically symmetric black-hole background,''
  Lett.\ Nuovo Cim.\  {\bf 3S2}, 211 (1972)
  [Lett.\ Nuovo Cim.\  {\bf 3}, 211 (1972)].
  doi:10.1007/BF02772872


 \bibitem{edestr} A.~Edelman and G.~Strang, ``Pascal matrices,'' Amer.\ Math.\ Monthly {\bf 111}, 189 (2004).


\bibitem{Newman:1968uj} 
  E.~T.~Newman and R.~Penrose,
  ``New conservation laws for zero rest-mass fields in asymptotically flat space-time,''
  Proc.\ Roy.\ Soc.\ Lond.\ A {\bf 305}, 175 (1968).
  doi:10.1098/rspa.1968.0112
  
  
  
 \end{thebibliography}
\end{document}